\documentclass[12pt]{iopart}

\renewcommand{\i}{{\mathrm i}}
\renewcommand{\d}{{\mathrm d}}
\def\bea{\begin{eqnarray}}
\def\eea{\end{eqnarray}}
\def\tsum{\mathop{\textstyle \sum }}
\def\tprod{\mathop{\textstyle \prod }}
\def\binom#1#2{{#1 \choose #2}}
\eqnobysec

\begin{document}
\jl{1}

\title[Coupling coefficients of SO($n$) and integrals]{Coupling 
coefficients of SO($n$) and integrals involving Jacobi and Gegenbauer 
polynomials}

\author{Sigitas Ali\v sauskas}

\address{Institute of Theoretical Physics and Astronomy of Vilnius
University, \\
A. Go\v stauto 12, Vilnius 2600, Lithuania}

\begin{abstract}
The expressions of the coupling coefficients ($3j$-symbols) for the most
degenerate (symmetric) representations of the orthogonal groups SO($n$) 
in a canonical basis (with SO($n$) restricted to SO($n-1$)) and 
different semicanonical or tree bases [with SO($n$) restricted to 
SO$(n^{\prime })\times $SO$(n^{\prime \prime })$, 
$n^{\prime }+n^{\prime \prime }=n$] are considered, respectively, in 
context of the integrals involving triplets of the Gegenbauer and the 
Jacobi polynomials. Since the directly derived triple-hypergeometric 
series do not reveal the apparent triangle conditions of the 3j-symbols, 
they are rearranged, using their relation with the semistretched 
isofactors of the second kind for the complementary chain 
Sp(4)$\supset $SU(2)$\times $SU(2) and analogy with the stretched $9j$ 
coefficients of SU(2), into formulae with more rich limits for summation 
intervals and obvious triangle conditions. The isofactors of class-one 
representations of the orthogonal groups or class-two representations of 
the unitary groups (and, of course, the related integrals involving 
triplets of the Gegenbauer and the Jacobi polynomials) turn into the 
double sums in the cases of the canonical SO($n$)$\supset $SO($n-1$) or 
U($n$)$\supset $U($n-1$) and semicanonical 
SO$(n)\!\supset $SO$(n-2)\times $SO(2) chains, as well as into the 
$_4F_3(1)$ series under more specific conditions. Some ambiguities of the 
phase choice of the complementary group approach are adjusted, as well 
as the problems with alternating sign parameter of SO(2) representations 
in the SO(3)$\supset $SO(2) and SO$(n)\!\supset $SO$(n-2)\times $SO(2) 
chains.
\end{abstract}


\maketitle

\section{Introduction}

The coupling (Clebsch--Gordan) coefficients and $3j$-symbols (Wigner 
coefficients) of the orthogonal groups SO($n$), together with their 
isoscalar factors (isofactors), maintain great importance in many fields 
of theoretical physics such as atomic, nuclear and statistical physics. 
The representation functions in terms of the Gegenbauer (ultraspherical) 
polynomials are well known for the symmetric (also called most degenerate 
or class-one) irreducible representations (irreps) of SO$(n)$ in the 
spherical coordinates (Vilenkin \cite{Vi65a}) on the unit sphere 
$S_{n-1}$. In particular, the explicit Clebsch--Gordan (CG) coefficients 
and isofactors of SO($n$) in the canonical basis for all three irreps 
symmetric were considered by Gavrilik \cite{Ga73}, Kildyushov and 
Kuznetsov \cite{KK73} (see also \cite{KMSS92}) and Junker \cite{Ju93}, 
using the direct \cite{Ga73,Ju93} or rather complicated indirect 
\cite{KK73,KMSS92} integration procedures.

Norvai\v sas and Ali\v sauskas \cite{NAl74a} also derived triple-sum 
expressions for related isofactors of SO($n$) in the case of 
the canonical (labelled by the chain of groups SO$(n)\!\supset $SO($n-1$)) 
and semicanonical bases (labelled by irreps $l,l^{\prime },
l^{\prime \prime }$ of the group chains SO$(n)\!\supset $SO$(n^{\prime 
})\times $SO$(n^{\prime \prime })$, $n^{\prime }+n^{\prime \prime }=n$, 
in the polyspherical, or the tree type, coordinates 
\cite{Vi65a,KMSS92,Vi65b}), exploiting the transition matrices 
\cite{AlV72} (also cf \cite{K97}) between the bases, labelled by the 
unitary and orthogonal subgroups in the symmetrical irreducible spaces of 
the U$(n)$ group. They observed \cite{NAl74a,NAl74b} that isofactors for 
the group chain SO$(n)\!\supset $SO$(n^{\prime })\times $SO$(n^{\prime 
\prime })$ for coupling of the states of symmetric irreps $l_1,l_2$ are 
the analytical continuation of the isofactors for the chain
Sp(4)$\supset $SU(2)$\times $SU(2), 
\bea
\fl \left[ \begin{array}{ccc}
l_1 & l_2 & [L_1L_2] \\ 
l_1^{\prime },l_1^{\prime \prime } & l_2^{\prime },l_2^{\prime
\prime } & \gamma [L_1^{\prime }L_2^{\prime }][L_1^{\prime \prime
}L_2^{\prime \prime }]
\end{array} 
\right] _{(n:n^{\prime }n^{\prime \prime })}  \nonumber \\
\lo= (-1)^{\phi } \left[ \begin{array}{ccc}
\!\left\langle \frac{-2L_2^{\prime }-n^{\prime }}{4}\frac{-2L_1^{\prime
}-n^{\prime }}{4}\right\rangle \! & \!\left\langle \frac{-2L_2^{\prime 
\prime }-n^{\prime \prime }}{4}\frac{-2L_1^{\prime \prime }-n^{\prime 
\prime }}{4}\right\rangle \! & \!\left\langle \frac{-2L_2-n}{4}\frac{-2L_1
-n}{4}\right\rangle ^{\!\gamma }\! \\ 
\frac{-2l_1^{\prime }-n^{\prime }}{4},\frac{-2l_2^{\prime }-n^{\prime %
}}{4} & \frac{-2l_1^{\prime \prime }-n^{\prime \prime }}{4},\frac{%
-2l_2^{\prime \prime }-n^{\prime \prime }}{4} & \frac{-2l_1-n}{4},
\frac{-2l_2-n}{4}
\end{array} \right]  \label{anct}
\eea
(i.e.\ they coincide, up to phase factor $(-1)^{\phi }$, with the 
isofactors for the non-compact complementary group [11--13] %\cite{MQ70,MQ71,Q73} 
chain Sp(4,$R$)$\supset $Sp(2,$R$)$\times $Sp(2,$R$) in the case for the 
discrete series of irreps). Particularly, in special multiplicity-free 
case (for $L_2=L_2^{\prime }=L_2^{\prime \prime }=0$, when the 
label $\gamma $ is absent), isofactors of SO$(n)\!\supset $SO($n-1$) 
correspond to the semistretched isofactors of the second kind 
\cite{AlJ71} of Sp(4)$\supset $SU(2)$\times $SU(2) (see also 
\cite{Al83,Al87}). 

However, neither expressions derived by means of the direct integration
\cite{Ga73,Ju93}, nor expressions derived by the re-expansion of states 
of the group chains \cite{NAl74a,NAl74b} reveal the apparent triangle 
conditions of the 3j-symbols in these triple-sum series. Only the 
substitution group technique of the Sp(4) or SO(5) group \cite{AlJ69}, 
used together with an analytical continuation procedure, enabled an 
indication \cite{NAl74a,NAl74b,Al83} of the transformation of the initial 
triple-sum expressions of \cite{NAl74a} into other forms, more 
convenient in the cases close to the stretched ones (e.g.\ for small 
values of $l_1+l_2-l_3$, where $l_3=L_1$) and turning into the double 
sums for the canonical basis, the SO$(n)\!\supset $SO$(n-2)\times $SO(2) 
chain and other cases with specified parameters 
$l_1^{\prime \prime }+l_2^{\prime \prime }-l_3^{\prime \prime }
=0$ (where $l_3^{\prime \prime }=L_1^{\prime \prime }$). 
More specified isofactors of SO$(n)\!\supset $SO($n-1$) \cite{Al87} are 
related to $6j$ coefficients of SU(2) (with some parameters being 
multiple of 1/4 in the case where $n$ is odd).

Unfortunately, the empirical phase choices of isofactors in 
\cite{NAl74a,NAl74b,Al83,Al87}\footnote{Note that the phase factor 
$(-1)^{(g-e)/2}$ (where $g\geq e$) should be omitted on the right-hand 
side of expression (5.7) of \cite{Al87} for recoupling ($6l$) 
coefficients of SO($n$), in contrast with (5.5) of the same paper.} were 
not correlated with the basis states (cf.\ \cite{Vi65a,KMSS92,IPSW01}) in 
terms of the Gegenbauer and the Jacobi polynomials. Some aspects of the 
isofactor symmetry problem were also left untouched, e.g.\ the problem of 
the sign change for irreps $m$ of the SO(2) subgroups (which was not 
revealed in \cite{Vi65a,KMSS92,IPSW01} for the states of 
SO(3)$\supset $SO(2) and SO$(n)\!\supset $SO$(n-2)\times $SO(2) either), 
as well as the indefiniteness of the type $(2l^{\prime \prime }+n^{\prime 
\prime }-2)(n^{\prime \prime }-4)!!$ for $l^{\prime \prime }=0$, 
$n^{\prime \prime }=2$ in numerator or denominator.

Presentation of the unambiguous proof of the most preferable and 
consistent expressions for the $3j$-symbols of the orthogonal SO($n$) and 
unitary U($n$) groups for decomposition of the factorized ultraspherical 
and polyspherical harmonics (i.e.\ for coupling of three most degenerate 
irreps into scalar representation in the cases of the canonical and 
semicanonical bases) is the main intention of this paper (cf [2--6,~16]) 
%\cite{Ga73,KK73,KMSS92,Ju93,NAl74a,Al87}; 
and is impossible without including a comprehensive review of some 
adjusted previous results \cite{Ju93,NAl74a,AlJ71,Al83}, since some 
references [2--4,~6,~15] %\cite{Ga73,KK73,KMSS92,NAl74a,Al83} 
may be not easily accessible nor free from misprints. However, the main 
goal of this paper is a strict {\it ab initio} rearrangement of the most 
symmetric (although banal) finite triple-sum series of the 
hypergeometric-type in the expressions of the definite integrals 
involving triplets of the multiplied Gegenbauer and Jacobi polynomials 
into less symmetric but more convenient triple (\ref{iJpd}), double 
(\ref{iGpc}), or single (\ref{iGpr}) sum series with summation intervals 
depending on the triangular conditions of the corresponding $3j$-symbols. 

The related triple-hypergeometric series, appearing in the expressions 
for the semistretched isoscalar factors of the second kind of the chain 
Sp(4)$\supset $SU(2)$\times $SU(2), are considered in section 2, together 
with their {\it ab initio} rearrangement using the different expressions 
\cite{Al00} for the stretched $9j$ coefficients of SU(2). (These triple 
sum series may be treated as extensions of the double-hypergeometric 
series of Kamp\'{e} de F\'{e}riet \cite{K-F21} type, e.g. considered by 
Lievens and Van der Jeugt \cite{LV-J01}.)

Well known special integral involving triplet of the Jacobi polynomials 
$P_k^{(\alpha ,\beta )}(x)$ \cite{BE53,AS65} in terms of the
Clebsch--Gordan coefficients of SU(2) (cf \cite{Vi65a}) 
\bea
\fl \frac 12\int\limits_{-1}^1\d x\left( \frac{1+x}{2}\right)
^{(\beta _1+\beta _2+\beta _3)/2}\left( \frac{1-x}{2}\right) 
^{(\alpha _1+\alpha _2+\alpha _3)/2}\prod_{a=1}^{3}P_{k_a}
^{(\alpha _a,\beta _a)}(x)  \nonumber \\
\lo= \left( \frac{1}{2l_3+1}\prod_{a=1}^{3}\frac{(k_a+\alpha _a)!
(k_a+\beta _a)!}{k_a!(k_a+\alpha _a+\beta _a)!}\right) ^{1/2} 
C_{m_1 m_2 m_3}^{l_1 l_2 l_3}C_{n_1 n_2 n_3}^{l_1 l_2 l_3},  \label{iJps}
\eea
may be derived only in frames of the angular momentum theory [24--26], 
%\cite{JB77,VMK88,BL81}
when 
\[
l_a=k_a+\case 12(\alpha _a+\beta _a),\qquad m_a=\case 12
(\alpha _a+\beta _a),\qquad n_a=\case 12(\beta _a-\alpha _a) 
\]
and 
\begin{eqnarray*}
\alpha _a=m_a-n_a,\qquad \beta _a=m_a+n_a,\qquad k_a=l_a-m_a; \\
\alpha _3=\alpha _1+\alpha _2,\qquad \beta _3=\beta _1+\beta _2 
\end{eqnarray*}
are non-negative integers. Unfortunately, quite an elaborate expansion 
\cite{KK73,KMSS92} of two multiplied Jacobi or Gegenbauer polynomials in 
terms the third Jacobi or Gegenbauer polynomial in frames of (\ref{iJps}) 
gives rather complicated multiple-sum expressions for integrals involving 
the ultraspherical or polyspherical functions in the generic SO($n$) or 
U($n$) case. In section 3, the definite integrals involving triplets of 
the multiplied unrestrained Jacobi and Gegenbauer polynomials at first 
are expressed using the direct (cf \cite{Ga73,Ju93}) integration 
procedure as the triple-sums in terms of beta and gamma functions. Later 
they are rearranged to more convenient forms, with fewer number of sums, 
or at least, with richer structure of the summation intervals 
(responding to the triangular conditions of the coupling coefficients) 
and better possibilities of summation (especially, under definite 
restrictions or for some coinciding parameters).

In section 4, some normalization and phase choice peculiarities of 
the canonical basis states and matrix elements of the symmetric 
(class-one) irreducible representations of SO($n$) are discussed. 
Then we consider the corresponding expressions of $3j$-symbols and 
Clebsch--Gordan coefficients of SO($n$), factorized in terms of integrals 
involving triplets of the Gegenbauer polynomials (preferable in 
comparison with results of \cite{Ju93}) and extreme (summable) 
$3j$-symbols, together with the alternative phase systems.

In section 5, the semicanonical basis states and matrix elements of the
symmetric (class-one) irreducible representations of SO$(n)$ for 
restriction SO$(n)\!\supset $SO$(n^{\prime })\times $SO$(n^{\prime 
\prime })$ ($n^{\prime }+n^{\prime \prime }=n$) are discussed. The 
corresponding factorized $3j$-symbols and Clebsch--Gordan coefficients, 
expressed in terms of integrals involving triplets of the Jacobi 
polynomials and extreme $3j$-symbols, are considered, together with 
special approach to the $n^{\prime \prime }=2$ and 
$n^{\prime }=n^{\prime \prime }$ cases.

The spherical functions for the canonical chain of unitary groups 
U$(n)\!\supset $U$(n-1)\times $U(1)$\supset \cdot \cdot \cdot \supset $%
U$(2)\times $U$(1)\!\supset $U$(1)$ correspond to the matrix elements of 
the class-two (mixed tensor) representations of U($n$), which include 
scalar of subgroup U($n-1$) (see \cite{KMSS92}). The factorized 
$3j$-symbols of U($n$), related in this case to isofactors of 
SO$(2n)\!\supset $SO$(2n-2)\times $SO(2) and expressed in terms of 
special integrals involving triplets of the Jacobi polynomials, are
considered in section 6.

In Appendix A, some special cases of the triple-sum series, used in 
section 2, are given as turning into the double- or single sum series for 
some coinciding values of parameters. An elementary proof of our main 
expression (\ref{iJpd}) for integrals is added as Appendix B.

\section{Semistretched isoscalar factors of the second kind of Sp(4) and
their rearrangement}

Here the irreducible representations of Sp(4) will be denoted by 
$\langle K\Lambda \rangle $, where the pairs of parameters 
$K=I_{\max },\Lambda =J_{\max }$ correspond to the maximal values of 
irreps $I$ and $J$ of the maximal subgroup SU(2)$\times $SU(2) (see 
\cite{AlJ71,AlJ69,H65}) and to the irreps of SO(5) with the highest 
weight $[L_1L_2]=[K+\Lambda ,K-\Lambda ]$. Below we consider the triple 
series shown in \cite{AlJ71}, where the following expression for the 
semistretched isoscalar factors of the second kind (with the coupled and 
resulting irrep parameters matching condition $K_1+K_2=K$) for the basis 
labelled by the chain Sp(4)$\supset $SU(2)$\times $SU(2) was derived: 
\bea
\fl \left[ \begin{array}{ccc}
\left\langle K_1\Lambda _1\right\rangle & \left\langle K_2\Lambda
_2\right\rangle & \left\langle K_1+K_2,\Lambda \right\rangle \\ 
I_1J_1 & I_2J_2 & I\,J
\end{array} \right]  \nonumber \\
\lo= (-1)^{\Lambda _1+\Lambda _2-\Lambda }\left[
(2I_1+1)(2J_1+1)(2I_2+1)(2J_2+1)(2\Lambda +1)\right] ^{1/2} 
\nonumber \\
\times \left[ \frac{\prod_{a=1}^{2}(2K_a-2\Lambda _a)!(2K_a+1)!
(2K_a+2\Lambda _a+2)!}{(2K_1+2K_2-2\Lambda )!(2K_1+2K_2+1)!
(2K_1+2K_2+2\Lambda +2)!}\right] ^{1/2}  \nonumber \\
\times \left[ \begin{array}{c}
K_1 \\ K_2 \\ K_1+K_2
\end{array} \left| \begin{array}{lll}
\Lambda _1 & I_1 & J_1 \\ 
\Lambda _2 & I_2 & J_2 \\ 
\Lambda & I & J
\end{array} \right. \right] .  \label{ssisf5}
\eea
Here $11j$ coefficient \cite{AlJ71} 
\bea
\fl \left[ \begin{array}{c}
K_1 \\ K_2 \\ K_1+K_2
\end{array} \left| \begin{array}{lll}
\Lambda _1 & I_1 & J_1 \\ 
\Lambda _2 & I_2 & J_2 \\ 
\Lambda & I & J
\end{array} \right. \right] =\frac{E(K_1+K_2+\Lambda ,I,J)}{%
\prod_{a=1}^{2}E(K_a+\Lambda _a,I_a,J_a)\nabla (K_a-\Lambda _a,
I_a,J_a)}  \nonumber \\
\times \frac{\nabla (K_1+K_2-\Lambda ,I,J)}{\nabla (II_1I_2)
\nabla (JJ_1J_2)\nabla (\Lambda \Lambda _1\Lambda _2)}\;
\widetilde{\bf S}\left[ \begin{array}{cccc}
K_1 & I_1 & J_1 & \Lambda _1 \\ 
K_2 & I_2 & J_2 & \Lambda _2 \\ 
K_1+K_2 & I & J & \Lambda
\end{array} \right]  \label{c11j5}
\eea
(which does not belong to $3nj$ coefficients of the angular momentum 
theory) is expressed in terms of the triple-sum 
$\widetilde{\bf S}[ \cdot \cdot \cdot ]$ and is invariant under 
permutations of the three right-hand columns, when the transposition of 
the first two rows gives the phase factor 
\begin{equation}
(-1)^{I_1+I_2-I+J_1+J_2-J+\Lambda _1+\Lambda _2-\Lambda }.
\label{ants}
\end{equation}
In (\ref{c11j5}) and further we use the notation 
\numparts \bea
\fl \nabla (abc)=\left[ \frac{(a+b-c)!(a-b+c)!(a+b+c+1)!}{(b+c-a)!}
\right] ^{1/2}  \label{nabla} \\
\lo= \left[ \frac{\Gamma (a+b-c+1)\Gamma (a-b+c+1)\Gamma (a+b+c+2)}{%
\Gamma (b+c-a+1)}\right] ^{1/2}  \label{nablg}
\eea \endnumparts
and
\begin{equation}
\fl E(abc)=\left[ (a-b-c)!(a-b+c+1)!(a+b-c+1)!(a+b+c+2)!\right] ^{1/2}.
\label{eabc}
\end{equation}

Now we present different expressions of the triple sum $\widetilde{%
{\bf S}}[\cdot \cdot \cdot ]$ that appear in (\ref{c11j5}): 
\numparts \bea
\fl \widetilde{{\bf S}}\left[ 
\begin{array}{cccc}
K_1 & j_1^1 & j_1^2 & j_1^3 \\ 
K_2 & j_2^1 & j_2^2 & j_2^3 \\ 
K_1+K_2 & j^1 & j^2 & j^3
\end{array} \right] =\sum_{x_1,x_2,x_3}\binom{j^1+j^2+j^3-K_1-K_2}{%
\sum_{a=1}^{3}(j_1^a-x_a)-K_1}  \nonumber \\
\times \prod_{a=1}^{3}\frac{(-1)^{x_a}(2j_1^a-x_a)!(j^a-j_1^a
+j_2^a+x_a)!}{x_a!(j_1^a+j_2^a-j^a-x_a)!}  \label{s11ja} \\
\lo= (j_1^1-j_2^1+j^1)!(j_1^1+j_2^1+j^1+1)!(j^2-j_1^2+j_2^2)!
(j_1^2+j_2^2+j^2+1)! \nonumber \\
\times \sum_{x_3,u,v}\frac{(-1)^{j_1^1+j_2^1-j^1+x_3+u+v}(2j_1^3-x_3)!
(j^3-j_1^3+j_2^3+x_3)!}{x_3!(j_1^3\!+\!j_2^3\!-\!j^3\!-\!x_3)!v!
(j_1^1\!+\!j_2^1\!-\!j^1\!-\!v)!(j_1^1\!+\!j_2^1\!+\!j^1\!-\!v\!+\!1)!}  
\nonumber \\
\times \frac{(2j_2^1-v)!(2j_1^2-u)!}{(j_2^1+j_2^2+j^3-j_1^3
-K_2+x_3-v)!u!(j_1^2+j_2^2-j^2-u)!}  \nonumber \\
\times \frac{(j_1^1+j_2^1+j_1^2+j_2^2+j^3-K_1-K_2-u-v)!}{(j_1^2+j_2^2
+j^2-u+1)!(j_1^1+j_1^2+j_1^3-K_1-x_3-u)!}  \label{s11jb} \\
\lo= (-1)^{K_1-j_1^1-j_1^2+j_1^3}\frac{(j_1^3-j_2^3+j^3)!}{(j_1^3+j_2^3
-j^3)!}\prod_{a=1}^{2}(j^a-j_1^a+j_2^a)!(j_1^a+j_2^a+j^a+1)!  \nonumber \\
\times \sum_{x_1,x_2,x_3}\prod_{a=1}^{2}\frac{(2j_1^a-x_a)!}{x_a!
(j_1^a+j_2^a-j^a-x_a)!(j_1^a+j_2^a+j^a-x_a+1)!}  \nonumber \\
\times \frac{(-1)^{x_3}(2j_1^3-x_3)!(j^3-j_1^3+j_2^3+x_3)!}{x_3!(j_1^3
-j_2^3+j^3-x_3)!}\binom{K_2-j_2^1-j_2^2+j_2^3}{\sum_{a=1}^{3}(j_1^a-x_a)
-K_1}.  \label{s11jc}
\eea \endnumparts
For rearrangement of (\ref{s11ja}) into (\ref{s11jc}) we used the 
different expressions of the stretched $9j$ coefficients \cite{Al00}. We 
transformed the double sum over $x_1,x_2$ in (\ref{s11ja}) into the 
sum over $u,v$ in (\ref{s11jb}) using relation 
(C1$a$)--(C1$c$) of \cite{Al00} and later the double sum over 
$v,x_3$ in (\ref{s11jb}) into the sum over $x_1,x_3$ in 
(\ref{s11jc}) using relation (C1$f$)--(C1$b$) of \cite{Al00} and 
replacing $u$ by $x_2$. (The related transformations for the double 
hypergeometric series of Kamp\'{e} de F\'{e}riet-type \cite{K-F21} are 
also considered by Lievens and Van der Jeugt \cite{LV-J01}.)

We see that all three summation parameters are restricted by 
$j_1^1+j_1^2+j_1^3-K_1$, or by $j_2^1+j_2^2+j_2^3-K_2$ in (\ref{s11ja}), 
as well as by $j_1^1+j_1^2+j_1^3-K_1$, or by $K_1+K_2-j^1-j^2+j^3$ in 
(\ref{s11jc}) [respectively, by $j_2^1+j_2^2+j_2^3-K_2$, or by 
$K_1+K_2+j^i-j^k-j^l$, ($i,k,l=1,2,3$) in the different versions 
of (\ref{s11jc}), related by symmetries]. In other cases the interval for 
the linear combination of summation parameters $x_1+x_2+x_3$ is 
restricted by $j^1+j^2+j^3-K_1-K_2$ in (\ref{s11ja}), as well as by 
$K_2-j_2^1-j_2^2+j_2^3$ in (\ref{s11jc}) [respectively, by 
$K_a+j_a^i-j_a^l+j_a^k$, ($i,k,l=1,2,3;\;a=1,2$) in the different 
versions of (\ref{s11jc}), related by symmetries]. Hence, there are 
five possibilities of the completely summable expressions for 
$\widetilde{\bf S}[\cdot \cdot \cdot ]$ and seven cases when they turn 
into double sums, dissimilar with nine cases, related to the stretched 
$9j$ coefficients \cite{Al00,JB77}.

For our further applications, it is more convenient to write relations 
(\ref{s11ja})--(\ref{s11jc}) (divided by $\prod_{a=1}^{3}(j^a-j_1^a
+j_2^a)!(j^a+j_1^a-j_2^a)!$) in another parametrization: 
\numparts \bea
\fl \widetilde{\cal S}\left[ 
\begin{array}{cccc}
\alpha _0,\beta _0 & \alpha _1,\beta _1 & \alpha _2,\beta _2 & 
\alpha _3,\beta _3 \\ 
& k_1 & k_2 & k_3
\end{array}
\right]  \nonumber \\
\lo= \sum_{z_1,z_2,z_3}\binom{\frac 12(\alpha _0+\beta
_0)-\sum_{a=1}^{3}[k_a+\frac 12(\alpha _a+\beta _a)]-2}{%
\frac 12\beta _0-\sum_{a=1}^{3}(\frac 12\beta _a+z_a)-1}  
\nonumber \\
\times \prod_{a=1}^{3}\frac{(-1)^{z_a}(-k_a-\alpha
_a)_{z_a}(-k_a-\beta _a)_{k_a-z_a}}{z_a!(k_a-z_a)!}
\label{s11pa} \\
\lo= \binom{-(2k_i\!+\!\alpha _i\!+\!\beta _i\!+\!2)}{-(k_i+\alpha _i+1)}%
\sum_{z_1,z_2,z_3}(-1)^{p_i-p_i^{\prime \prime }+z_1+z_2
+z_3}\binom{p_i^{\prime \prime }}{p_i\!-\!z_1\!-\!z_2\!-\!z_3}  
\nonumber \\
\times \frac{(k_i+1)_{z_i}(k_i+\beta _i+1)_{z_i}}{z_i!(2k_i
+\alpha _i+\beta _i+2)_{z_i}}\prod_{a\neq i}\frac{(-k_a\!-\!\beta _a)_{%
z_a}(k_a\!+\!\alpha _a\!+\!\beta _a\!+\!1)_{k_a-z_a}}{z_a!(k_a-z_a)!}.  
\label{s11pb}
\eea \endnumparts
We use the Pochhammer symbols 
\[
(c)_n=\prod_{k=0}^{n-1}(c+k)=\frac{\Gamma (c+n)}{\Gamma (c)}
\]
and binomial coefficients, which arguments are the non-negative integers.
Here 11 parameters of (\ref{s11ja}) and (\ref{s11jc}) are replaced by 
\begin{eqnarray*}
\fl k_1=I_1+I_2-I,\qquad k_2=J_1+J_2-J,\qquad k_3=
\Lambda _1+\Lambda _2-\Lambda ; \\
\fl \alpha _0=-2K_2-1,\qquad \alpha _1=-2I_2-1,\qquad \alpha _2=-2J_2-1,
\qquad \alpha _3=-2\Lambda _2-1; \\
\fl \beta _0=-2K_1-1,\qquad \beta _1=-2I_1-1,\qquad \beta _2=-2J_1-1,
\qquad \beta _3=-2\Lambda _1-1,
\end{eqnarray*}
with the non-negative integers 
\begin{eqnarray*}
p_i^{\prime }=\case 12(\beta _j+\beta _k-\beta _i-\beta _0),\qquad 
p_i^{\prime \prime }=\case 12(\alpha _j+\alpha _k-\alpha _i
-\alpha _0), \\
p_i=k_j+k_{k}-k_i+p_i^{\prime }+p_i^{\prime \prime }\qquad 
(i,j,k=1,2,3)
\end{eqnarray*}
and arguments of binomial coefficients, although parameters $\alpha _j$ 
and $\beta _j$ ($j=0,1,2,3$) here are negative integers. Actually, 
expression (\ref{s11pb}) may be written in three versions.

\section{Integrals involving triplets of Jacobi and Gegenbauer 
polynomials}

It is convenient for our purposes to use the two following expressions 
for the Jacobi polynomials (cf (16) of section 10.8 of \cite{BE53} and 
chapter 22 of \cite{AS65}): 
\numparts \bea
\fl P_k^{(\alpha ,\beta )}(x)=2^{-k}\sum_{m}\frac{(-k-\alpha )_m(-k
-\beta )_{k-m}}{m!(k-m)!}(-1)^{m}(1+x)^{m}(1-x)^{k-m}  \label{dJpa} \\
\lo= (-1)^{k}\sum_{m}\frac{(-k-\alpha )_m(k+\alpha +\beta +1)_{k-m}}{m!
(k-m)!}\left( \frac{1-x}{2}\right) ^{k-m},  \label{dJpb}
\eea \endnumparts
where $\alpha >-1,\beta >-1$.

We introduce the following expressions for the integrals involving the 
product of three Jacobi polynomials $P_{k_1}^{(\alpha _1,\beta _1)}(x)$, 
$P_{k_2}^{(\alpha _2,\beta _2)}(x)$ and $P_{k_3}^{(\alpha _3,\beta _3)}
(x)$ with a measure dependent on $\alpha _0>-1,\beta _0>-1$ and integers 
$\alpha _a-\alpha _0\geq 0$, $\beta _a-\beta _0\geq 0$ ($a=1,2,3$): 
\numparts \bea
\fl \frac 12\int\limits_{-1}^1\d x\left( \frac{1+x}{2}\right)
^{(\beta _1+\beta _2+\beta _3-\beta _0)/2}\left( \frac{1-x}{2}%
\right) ^{(\alpha _1+\alpha _2+\alpha _3-\alpha _0)/2}
\prod_{a=1}^{3}P_{k_a}^{(\alpha _a,\beta _a)}(x)  \nonumber \\
\lo= \widetilde{\cal I}\left[ 
\begin{array}{cccc}
\alpha _0,\beta _0 & \alpha _1,\beta _1 & 
\alpha _2,\beta _2 & \alpha _3,\beta _3 \\ 
& k_1 & k_2 & k_3
\end{array}
\right]  \label{iJp}  \\
\lo= (-1)^{k_1+k_2+k_3}\widetilde{\cal I}\left[ 
\begin{array}{cccc}
\beta _0,\alpha _0 & \beta _1,\alpha _1 & 
\beta _2,\alpha _2 & \beta _3,\alpha _3 \\ 
& k_1 & k_2 & k_3
\end{array} \right]  \label{iJpt} \\
\lo= \sum_{z_1,z_2,z_3}{\rm B}\left( 1\!-\!\case 12\beta _0\!+\!%
\tsum_{a=1}^{3}(\case 12\beta _a\!+\!z_a),1\!-\!\case 12\alpha _0\!+\!%
\tsum_{a=1}^{3}(\case 12\alpha _a\!+\!k_a\!-\!z_a)\right)  \nonumber \\
\times \prod_{a=1}^{3}\frac{(-1)^{z_a}(-k_a-\alpha _a)_{z_a}(-k_a
-\beta _a)_{k_a-z_a}}{z_a!(k_a-z_a)!}  \label{iJpb} \\
\lo= \sum_{z_1,z_2,z_3}{\rm B}\left( 1-\case 12\beta _0+\case 12
\tsum_{a=1}^{3}\beta _a,1-\case 12\alpha _0+\tsum_{a=1}^{3}(\case 12
\alpha _a+k_a-z_a)\right)  \nonumber \\
\times \prod_{a=1}^{3}\frac{(-1)^{k_a}(-k_a-\alpha _a)_{z_a}(k_a+\alpha _a
+\beta _a+1)_{k_a-z_a}}{z_a!(k_a-z_a)!}  \label{iJpc} \\
\lo= {\rm B}( k_i+\alpha _i+1,k_i+\beta _i+1)\sum_{z_1,z_2,z_3}(-1)^{p_i
-p_i^{\prime \prime }+z_1+z_2+z_3}  \nonumber \\
\times \binom{p_i^{\prime \prime }}{p_i-z_1-z_2-z_3}\frac{(k_i+1)_{z_i}
(k_i+\beta _i+1)_{z_i}}{z_i!(2k_i+\alpha _i+\beta _i+2)_{z_i}} 
\nonumber \\
\times \prod_{a=j,k;a\neq i}\frac{(-k_a-\beta _a)_{z_a}(k_a+\alpha _a
+\beta _a+1)_{k_a-z_a}}{z_a!(k_a-z_a)!},  \label{iJpd}
\eea \endnumparts
where the linear combinations (triangular conditions)
\begin{eqnarray*}
p_i^{\prime } =\case 12(\beta _j+\beta _k-\beta _i-\beta _0)\geq 0,
\qquad p_i^{\prime \prime }=\case 12(\alpha _j+\alpha _k-\alpha _i
-\alpha _0)\geq 0, \\
p_i=k_j+k_{k}-k_i+p_i^{\prime }+p_i^{\prime \prime }\geq 0\qquad 
(i,j,k=1,2,3)
\end{eqnarray*}
are integers. These integrals would otherwise vanish. Two first 
expressions (\ref{iJpb}) and (\ref{iJpc}) (including $(k_1+1)(k_2+1)
(k_3+1)$ terms each) are straightforward to derive using expressions 
(\ref{dJpa}) or (\ref{dJpb}) and definite integrals (see equation 
(6.2.1) of \cite{AS65}) in terms of beta functions ${\rm B}(x,y)=
\Gamma (x)\Gamma (y)/\Gamma (x+y)$. However, vanishing of integrals 
(\ref{iJp}) under spoiled triangular conditions is only seen directly in 
the estimated final expression (\ref{iJpd}), which cannot be derived in 
a similar manner as (\ref{iJpb}) and (\ref{iJpc}).
 
Although parameters $\alpha _a$, $\beta _a$ ($a=0,1,2,3$) accept the 
mutually excluding values in the sums $\widetilde{\cal S}[\cdot \cdot 
\cdot ]$ and $\widetilde{\cal I}[\cdot \cdot \cdot ]$, we only see the 
one-to-one correspondence of the analytical continuation between series 
(\ref{s11pa}) and (\ref{iJpb}), as well as between series (\ref{s11pb}) 
and (\ref{iJpd}), if the corresponding binomial coefficients of 
(\ref{s11pa}) and (\ref{s11pb}) depending on all negative (integer or 
half-integer) parameters are replaced in (\ref{iJpb}) and (\ref{iJpd}), 
respectively, by the beta functions. The possible zeros or poles of 
$\widetilde{\cal S}[\cdot \cdot \cdot ]$ for parameters of the definite 
binomial coefficients accepting negative integer (or half-integer) values 
may be disregarded, if the functions $\widetilde{\cal S}[\cdot \cdot 
\cdot ]\binom{-\alpha _0-\beta _0-2)}{-\alpha _0-1)}^{-1}$ and 
$\widetilde{\cal I}[\cdot \cdot \cdot ]{\rm B}^{-1}(\alpha _0+1,
\beta _0+1)$ are considered. Observing that the ratio of the binomial 
coefficients 
\[
\binom{-a-b-2}{-a-1}\binom{-c-d-2}{-c-1}^{-1} 
\]
with negative integers $a,b,c,d$ in equation (\ref{s11pa})--(\ref{s11pb}) 
turns into ratio of the beta functions 
\[
\frac{{\rm B}(a+1,b+1)}{{\rm B}(c+1,d+1)} 
\]
with parameters $a,b,c,d\geq -\frac 12$ in relation 
(\ref{iJpb})--(\ref{iJpd}), expression (\ref{iJpd}) for integral 
$\widetilde{\cal I}\left[ \cdot \cdot \cdot \right] $ is proved. An 
advantage of our new expression (\ref{iJpd}) is in the restriction of all 
three summation parameters $z_1+z_2+z_3$ by the triangular condition 
$p_i$, in contrast with (\ref{iJpb}) or (\ref{iJpc}). Alternatively, the 
linear combination of summation parameters $p_i-z_1-z_2-z_3\geq 0$ is 
restricted in addition by $p_i^{\prime \prime }$ (or by $p_i^{\prime }$, 
if some symmetry is applied) only in the $i$th version of (\ref{iJpd}). 
Hence, there are three cases when expressions for integral 
$\widetilde{\cal I}\left[ \cdot \cdot \cdot \right] $ are completely 
summable and six cases when they turn into double sums, in addition to 
the double sums which appear for $k_a=0$ ($a=1,2,3$). However, 
factorization of (\ref{iJpb})--(\ref{iJpd}) for $\alpha _0=\beta _0=0$, 
$\alpha _i=\alpha _j+\alpha _k$, $\beta _i=\beta _j+\beta _k$ into a
product of two CG coefficients of SU(2) is not straightforward to prove.

For $\alpha _0=0$ and $\alpha _3=\alpha _1+\alpha _2$, the integrals 
involving the product of three Jacobi polynomials (\ref{iJpd}) turn into 
the double sums (\ref{iJra}) or (\ref{iJrb})
\numparts \bea
\fl \widetilde{\cal I}\left[ \begin{array}{cccc}
0,\beta _0 & \alpha _1,\beta _1 & \alpha _2,\beta _2 & 
\alpha _1+\alpha _2,\beta _3 \\ 
& k_1 & k_2 & k_3
\end{array} \right] ={\rm B}(k_3+\alpha _1+\alpha _2+1,k_3+\beta _3+1) 
\nonumber \\
\times \sum_{z_1,z_2}\frac{(k_3+1)_{p_3-z_1-z_2}(k_3+\beta _3+1)_{p_3-z_1
-z_2}}{(p_3-z_1-z_2)!(2k_3+\alpha _1+\alpha _2+\beta _3+2)_{p_3-z_1-z_2}}  
\nonumber \\
\times \prod_{a=1}^{2}\frac{(-k_a-\beta _a)_{z_a}(k_a+\alpha _a+\beta
_a+1)_{k_a-z_a}}{z_a!(k_a-z_a)!}  \label{iJra} \\
\lo= {\rm B}\left( k_1+\alpha _1+1,k_1+\beta _1+1\right)   \nonumber \\
\times \sum_{z_2,z_3}\binom{k_2+\alpha _2}{z_2}\frac{(-1)^{\alpha _2-z_2}
(k_2+\alpha _2+\beta _2+1-z_2)_{k_2}(-k_3-\beta _3)_{z_3}}{z_3!(k_3-z_3)!
k_2!}  \nonumber \\
\times \frac{(k_3+\alpha _3+\beta _3+1)_{k_3-z_3}(k_1+1)_{p_1-z_2-z_3}(k_1
+\beta _1+1)_{p_1-z_2-z_3}}{(p_1-z_2-z_3)!(2k_1+\alpha _1+\beta _1+2)_{p_1
-z_2-z_3}},  \label{iJrb}
\eea \endnumparts
both related to the Kamp\'{e} de F\'{e}riet \cite{K-F21} functions 
$F_{2:1}^{2:2}$. It is evident that the triple series (\ref{iJpb}) and 
(\ref{iJpc}) with $\alpha _0=0$ may be also extended to the negative 
integer values of $\alpha _2$,
\bea
\fl \widetilde{\cal I}\left[ \begin{array}{cccc}
0,\beta _0 & \alpha _1,\beta _1 & \alpha _2,\beta _2 & 
\alpha _3,\beta _3 \\ 
& k_1 & k_2 & k_3
\end{array} \right]   \nonumber \\
\lo= (-1)^{\alpha _2}\frac{(k_2+\alpha _2)!(k_2+\beta _2)!}{%
k_2!(k_2+\alpha _2+\beta _2)!}\widetilde{\cal I}\left[ 
\begin{array}{cccc}
0,\beta _0 & \alpha _1,\beta _1 & -\alpha _2,\beta _2 & 
\alpha _3,\beta _3 \\ 
& k_1 & k_2+\alpha _2 & k_3
\end{array} \right] ,  \label{iJpe}
\eea
with invariant values of $p_1$ and $p_3$. Hence, using (\ref{iJra}) for 
the right-hand side of (\ref{iJpe}), the left-hand side of (\ref{iJpe}) 
may be expressed as the double sum for $\alpha _3=\alpha _1-\alpha _2$ 
and (\ref{iJrb}) may be derived after interchange of 
$k_1,\alpha _1,\beta _1$ and $k_3,\alpha _3,\beta _3$.

The Gegenbauer (ultraspherical) polynomial $C_k^{p}(\cos \theta )$ may be 
expressed as the finite series \cite{BE53,AS65}, or in terms of special 
Jacobi polynomial (cf \cite{BE53}) 
\numparts \bea
\fl C_k^{p}(\cos \theta )=\sum_{m=0}^{[k/2]}\frac{(-1)^{m}(p)_{k-m}}{m!
(k-2m)!}2^{k-2m}\cos ^{k-2m}\theta   \label{dGpa} \\
\lo= \frac{(2p)_k}{(p+1/2)_k}P_k^{(p-1/2,p-1/2)}(\cos \theta ),
\label{dGpJ}
\eea \endnumparts
where $[k/2]$ is an integer part of $k/2$ and (\ref{dGpJ}) includes 
almost twice as many terms as (\ref{dGpa}).

Now we may express the integrals involving the product of three 
Gegenbauer polynomials $C_{l_1-l_1^{\prime }}^{l_1^{\prime }+n/2-1}(x)$, 
$C_{l_2-l_2^{\prime }}^{l_2^{\prime }+n/2-1}(x)$ and 
$C_{l_3-l_3^{\prime }}^{l_3^{\prime }+n/2-1}(x)$ as follows: 
\numparts \bea
\fl \int\limits_0^{\pi }\d \theta (\sin \theta )^{l_1^{\prime
}+l_2^{\prime }+l_3^{\prime }+n-2}\prod_{i=1}^{3}C_{l_i-l_i^{%
\prime }}^{l_i^{\prime }+n/2-1}(\cos \theta )  \nonumber \\
\lo= \sum_{z_1,z_2,z_3}{\rm B}\left( \case 12(l_1^{\prime
}+l_2^{\prime }+l_3^{\prime }+n-1),\case 12+
\tsum_{a=1}^{3}[\case 12(l_a-l_a^{\prime })-z_a]\right)  \nonumber \\
\times \prod_{i=1}^{3}\frac{(-1)^{z_i}2^{l_i-l_i^{\prime
}-2z_i}(l_i^{\prime }+n/2-1)_{l_i-l_i^{\prime }-z_i}}{%
z_i!(l_i-l_i^{\prime }-2z_i)!}  \label{iGpa} \\
\lo= (-1)^{k_1+k_2+k_3}\prod_{a=1}^{3}\frac{(l_a^{\prime
}+n/2-1)_{(l_a-l_a^{\prime }+\delta _a)/2}}{(1/2)_{(l_a-l_a^{%
\prime }+\delta _a)/2}}  \nonumber \\
\times \widetilde{\cal I}\left[ 
\begin{array}{cccc}
\!-\frac 12,\frac{n-3}{2} & \!\delta _1\!-\!\frac 12,l_1^{\prime }\!%
+\!\frac{n-3}{2} & \!\delta _2\!-\!\frac 12,l_2^{\prime }\!+\!%
\frac{n-3}{2} & \!\delta _3\!-\!\frac 12,l_3^{\prime }\!+\!%
\frac{n-3}{2}\! \\ 
& \frac 12(l_1\!-\!l_1^{\prime }\!-\!\delta _1) & \frac 12(l_2\!%
-\!l_2^{\prime }\!-\!\delta _2) & \frac 12(l_3\!-\!l_3^{\prime }\!%
-\!\delta _3)
\end{array} \right]   \label{iGpb} \\
\lo= (-1)^{(l_j^{\prime }+l_k^{\prime }-l_i^{\prime })/2}{\rm B}%
\left( \case 12(l_i-l_i^{\prime }+\delta _i+1),\case 12%
(l_i+l_i^{\prime }-\delta _i+n-1)\right)   \nonumber \\
\times \prod_{a=1}^{3}\frac{(l_a^{\prime }+n/2-1)_{(l_a-l_a^{\prime
}+\delta _a)/2}}{(1/2)_{(l_a\!-\!l_a^{\prime }\!+\!\delta _a)/2}}%
\sum_{z_1,z_2,z_3}\binom{(\delta _j+\delta _k-\delta _i)/2}{(l_j\!%
+\!l_k\!-\!l_i)/2\!-\!z_1\!-\!z_2\!-\!z_3}  \nonumber \\
\times \prod_{a\neq i}\frac{\left( -(l_a\!+\!l_a^{\prime }\!-\!\delta
_a\!+\!n\!-\!3)/2\right) _{z_a}\left( (l_a\!+\!l_a^{\prime }\!+\!\delta
_a\!+\!n)/2\!-\!1\right) _{(l_a-l_a^{\prime }-\delta _a)/2-z_a}}{%
z_a!\left( (l_a-l_a^{\prime }-\delta _a)/2-z_a\right) !}  \nonumber \\
\times (-1)^{z_1+z_2+z_3}\frac{\left( (l_i-l_i^{\prime }-\delta _i)/2
+1\right) _{z_i}\left( (l_i+l_i^{\prime }-\delta _i+n-1)/2\right) 
_{z_i}}{z_i!(l_i+n/2)_{z_i}}  \label{iGpc} \\
\lo= \widetilde{\cal I}\left[ \begin{array}{cccc}
\overline{\alpha }_0,\overline{\alpha }_0 & 
l_1^{\prime }+\overline{\alpha }_0,l_1^{\prime }+\overline{\alpha }_0 & 
l_2^{\prime }+\overline{\alpha }_0,l_2^{\prime }+\overline{\alpha }_0 & 
l_3^{\prime }+\overline{\alpha }_0,l_3^{\prime }+\overline{\alpha }_0  \\
& l_1-l_1^{\prime } & l_2-l_2^{\prime } & l_3-l_3^{\prime }
\end{array} \right]   \nonumber \\
\times 2^{l_1^{\prime }+l_2^{\prime }+l_3^{\prime }+n-2}
\prod_{i=1}^{3}\frac{(2l_i^{\prime }+n-2)_{l_i-l_i^{\prime }}}{%
\left( l_i^{\prime }+(n-1)/2\right) _{l_i-l_i^{\prime }}},\qquad 
({\rm with}\quad \overline{\alpha }_{0}=\case{n-3}{2})  \label{iGpd} 
\eea \endnumparts
where $\frac 12(l_j+l_k-l_i)\geq 0$ and $\frac 12(l_j^{\prime }
+l_k^{\prime }-l_i^{\prime })\geq 0$ are integers. In accordance with 
(\ref{iGpc}), these integrals would otherwise vanish. Expressions 
(\ref{iGpa}) (cf \cite{Ju93}) and (\ref{iGpd}) (cf \cite{Ga73}) are 
derived directly (using definite integrals (6.2.1) of \cite{AS65} in 
terms of beta functions). Further (\ref{iGpa}) is recognized as 
consistent with a particular case of (\ref{iJpc})\footnote{This is the 
reason why (\ref{iJpc}) is introduced.} denoted by (\ref{iGpb}) and 
re-expressed, in accordance with (\ref{iJpc})--(\ref{iJpd}), in the most 
convenient form as (\ref{iGpc}), where $\delta _1,\delta _2,\delta _3=0$ 
or 1 (in fact either $\delta _1=\delta _2=\delta _3=0$, or $\delta _a=
\delta _b=1$, $\delta _c=0$) and $\frac 12(l_a-l_a^{\prime }-\delta _a)$ 
($a=1,2,3$) are integers.

Expression (\ref{iGpa}) includes $\frac 18\prod_{a=1}^{3}(l_a
-l_a^{\prime }-\delta _a+2)$ terms, when (\ref{iGpd}), used together 
with (\ref{iJpb}) or (\ref{iJpc}), each include $\prod_{a=1}^{3}(l_a
-l_a^{\prime }+1)$ terms; otherwise, the number of terms in the $i$th 
version of the most convenient formula (\ref{iGpc}) never exceeds 
\begin{equation}
\fl A_i=(p_i^{\prime \prime }+1)\min \left[ \case 12%
(p_i+1)(p_i-p_i^{\prime \prime }+2),\case 14\tprod_{a\neq
i}(l_a-l_a^{\prime }-\delta _a+2)\right] ,  \label{ntc}
\end{equation}
where $p_i^{\prime \prime }=\frac 12(\delta _j+\delta _k-\delta
_i)=0$ or 1, $p_i=\frac 12(l_j+l_k-l_i)$ is an integer and 
$i,j,k$ is a transposition of 1,2,3. This number of terms decreases in
comparison with (\ref{ntc}) in the intermediate region 
\[
\case 12\min (l_j-l_j^{\prime }-\delta _j,l_k-l_k^{\prime
}-\delta _k)<p_i<\case 12(l_j-l_j^{\prime }-\delta _j+l_k-l_k^{\prime
}-\delta _k). 
\]

Actually, expression (\ref{iGpc}) is related to the Kamp\'{e} de 
F\'{e}riet \cite{K-F21} function $F_{2:1}^{2:2}$ (for 
$p_i^{\prime \prime }=0$) or to the sum of two such functions (when 
$p_i^{\prime \prime }=1$). Hence, after comparing three different 
versions of (\ref{iGpc}), the rearrangement formulae of special 
Kamp\'{e} de F\'{e}riet functions $F_{2:1}^{2:2}$ can be derived.

Now we consider more specified integrals involving several Gegenbauer
polynomials. At first, using (\ref{iGpc}) with $i=3$ and $z_2=\delta
_2=0,\;z_3=\frac 12(l_1+l^{\prime }-l_3)-z_1$, we take special integral 
involving two multiplied Gegenbauer polynomials (where third trivial 
polynomial $C_0^{l^{\prime }+n/2-1}(x)=1$ may be inserted) in terms of 
the summable balanced (Saalsch\"{u}tzian) $_3F_2(1)$ series (cf 
\cite{Sl66,GR90}) and write: 
\numparts \bea
\fl \int\limits_0^{\pi }(\sin \theta )^{2l^{\prime }+n-2}C_{l_1
-l^{\prime }}^{l^{\prime }+n/2-1}(\cos \theta )C_{l^{\prime }-l^{\prime 
}}^{l^{\prime }+n/2-1}(\cos \theta )C_{l_3}^{n/2-1}(\cos \theta )
\d \theta  \nonumber \\
\lo= \int\limits_0^{\pi }(\sin \theta )^{2l^{\prime }+n-2}C_{l_1
-l^{\prime }}^{l^{\prime }+n/2-1}(\cos \theta )C_{l_3}^{n/2-1}(\cos 
\theta )\d \theta  \label{iGp2d} \\
\lo= \frac{(-1)^{(l_3-l_1+l^{\prime })/2}\pi \,l^{\prime }!(l_1
+l^{\prime }+n-3)!}{2^{2l^{\prime }+n-3}(l_1-l^{\prime })!
(J^{\prime }-l_1)!(J^{\prime }-l_3)!\Gamma (n/2-1)}  \nonumber \\
\times \frac{\Gamma (J^{\prime }-l^{\prime }+n/2-1)}{\Gamma (l^{\prime }
+n/2-1)\Gamma (J^{\prime }+n/2)},  \label{iGp2}
\eea \endnumparts
where $J^{\prime }=\case 12(l_1+l^{\prime }+l_3)$.

Using the expansion formula of two multiplied Gegenbauer polynomials as 
the zonal spherical functions 
\bea
\fl C_{l}^{p}(x)C_k^{p}(x)=\sum_{n=|l-k|}^{l+k}\frac{(n+p)\Gamma (g+2p)
\Gamma (g-n+p)}{\Gamma ^{2}(p)\Gamma (g+p+1)\Gamma (n+2p)\Gamma (g-n+1)}  
\nonumber \\
\times \frac{\Gamma (g-l+p)\Gamma (g-k+p)}{\Gamma (g-l+1)\Gamma (g-k+1)}%
C_n^{p}(x)\;  \label{eGpz}
\eea
in terms the third polynomial of the same type (cf \cite{Vi65a}), where 
$g=\frac 12(l+k+n)$ is an integer and $l+k-n$ is even, the special 
integral involving three Gegenbauer polynomials (with coinciding 
superscripts in two cases) also may be expanded in terms of integrals 
(\ref{iGp2}) and may be presented as follows:
\numparts \bea
\fl \int\limits_0^{\pi }(\sin \theta )^{2l^{\prime }+n-2}C_{l_1
-l^{\prime }}^{l^{\prime }+n/2-1}(\cos \theta )C_{l_2-l^{\prime }}^{%
l^{\prime }+n/2-1}(\cos \theta )C_{l_3}^{n/2-1}(\cos \theta )\d 
\theta  \nonumber \\
\lo= \frac{\pi l^{\prime }!}{2^{2l^{\prime }+n-3}\Gamma ^{3}(n/2-1)
\Gamma (l^{\prime }+n/2-1)}  \nonumber \\
\times \sum_{k=|l_1-l_2|+l^{\prime }}^{l_1+l_2-l^{\prime }}
\frac{(-1)^{(l_3+l^{\prime }-k)/2}(k+n/2-1)}{\nabla ^{2}\left( %
l^{\prime }/2,l_3/2+n/4-1,k/2+n/4-1\right) }  \nonumber \\
\times \frac{\nabla ^{2}\left( (l_1+l^{\prime }+n)/2-2,l_2/2+n/4-1,
k/2+n/4-1\right) }{\nabla ^{2}\left( (l_1-l^{\prime })/2,l_2/2+n/4-1,
k/2+n/4-1\right) }  \label{iGpk} \\
\lo= \frac{\pi \,l^{\prime }!\prod_{a=1}^{3}\Gamma (J-l_a+n/2-1)}{%
2^{2l^{\prime }+n-3}\Gamma (n/2-1)\Gamma (l^{\prime }+n/2-1)\Gamma 
(J+n/2)}  \nonumber \\
\times \sum_{u}\frac{(J+l^{\prime }+n-3-u)!}{u!(l^{\prime }-u)!(J-l_1-u)!
(J-l_2-u)!(J-l_3-l^{\prime }+u)!}  \nonumber \\
\times \frac{(-1)^{u}}{\Gamma (n/2-1+u)\Gamma (l^{\prime }+n/2-1-u)},  
\label{iGpr}
\eea \endnumparts
where $J=\case 12(l_1+l_2+l_3)$ and the gamma functions under
summation sign in the intermediate formula (\ref{iGpk}) (which is 
equivalent to (15) of \cite{HJu99}) are included into the asymmetric 
triangle coefficients (\ref{nablg}). Finally, the sum in (\ref{iGpk}) 
corresponds to the very well-poised $_7F_6(1)$ hypergeometric series 
(which may be rearranged using Watson's transformation formula (2.5.1) 
of \cite{GR90} or (6.10) of \cite{LB94} into balanced $_4F_3(1)$ 
hypergeometric series) or to the $6j$ coefficient 
\begin{equation}
\left\{ 
\begin{array}{ccc}
l^{\prime }+\frac 12n-2 & \frac 12(l_1+n)-2 & \frac 12l_1 \\ 
\frac 12l_3+\frac{1}{4}n-1 & \frac 12l_2+\frac{1}{4}n-1 & 
\frac 12l_2+\frac{1}{4}n-1
\end{array}
\right\}   \label{csR}
\end{equation}
with standard (for $n$ even) or multiples of 1/4 parameters, in 
accordance with expression (C3) of the $6j$ coefficient \cite{Al92} in 
terms of (\ref{nabla}). Using the most symmetric (Racah) expression 
\cite{JB77,VMK88} for (\ref{csR}), the final expression (\ref{iGpr}) with 
single sum is derived. Intervals of summation are restricted by 
$\min (l^{\prime },J-l_1,J-l_2,J-l_3)$ and, of course, (\ref{iGpr}) 
coincide with result of Vilenkin \cite{Vi65a} for $l^{\prime }=0$.

Comparing expansion (\ref{iGpc}) of integrals involving triplets of the 
Gegenbauer polynomials with (\ref{iGpd}), we may write expression for 
integrals involving triplets of special Jacobi polynomials, with mutually 
equal superscripts, 
\bea
\fl \widetilde{\cal I}\left[ \begin{array}{cccc}
\alpha _0,\alpha _0 & \alpha _1,\alpha _1 & 
\alpha _2,\alpha _2 & \alpha _3,\alpha _3 \\ 
& k_1 & k_2 & k_3
\end{array} \right]  \nonumber \\
\lo= \frac{[1+(-1)^{p_i}]\,{\rm B}(1/2,k_i+\alpha _i+1)}{2^{k_1+k_2+k_3
+\alpha _1+\alpha _2+\alpha _3-\alpha _0+2}(1/2)_{(k_j+\delta _j)/2}
(1/2)_{(k_{k}+\delta _k)/2}} \nonumber \\
\times \sum_{z_1,z_2,z_3}(-1)^{p_i^{\prime }+(k_j+\delta
_j+k_{k}+\delta _k)/2+z_1+z_2+z_3}\binom{(\delta _j+\delta
_k-\delta _i)/2}{p_i/2-z_1-z_2-z_3}  \nonumber \\
\times \prod_{a=j,k;a\neq i}\frac{\left( -k_a\!-\!\alpha _a\right)
_{(k_a+\delta _a)/2+z_a}\left( \alpha _a+(k_a+\delta
_a\!+\!1)/2\right) _{(k_a\!-\!\delta _a)/2\!-\!z_a}}{z_a!\left( (k_a
-\delta _a)/2-z_a\right) !}  \nonumber \\
\times \binom{(k_i-\delta _i)/2+z_i}{z_i}\frac{\left( \alpha _i+(k_i
-\delta _i)/2+1\right) _{z_i}}{(\alpha _i+k_i+3/2)_{z_i}}.  \label{iJpst}
\eea
Here 
\begin{eqnarray*}
p_i=k_j+k_{k}-k_i+p_i^{\prime }+p_i^{\prime \prime },\qquad 
\case 12(k_i-\delta _i),\qquad \delta _i=0\;{\rm or}\;1, \\
p_i^{\prime }=p_i^{\prime \prime }=\case 12(\alpha _j+\alpha
_k-\alpha _i-\alpha _0)\qquad (i,j,k=1,2,3)
\end{eqnarray*}
are non-negative integers.

Comparing expansion (\ref{iGpr}) of integrals involving more specified 
triplets of the Gegenbauer polynomials with (\ref{iGpd}), we may also 
write an expression for integrals involving triplets of special Jacobi 
polynomials, 
\bea
\fl \widetilde{\cal I}\left[ \begin{array}{cccc}
\alpha _0,\alpha _0 & \alpha _1,\alpha _1 & 
\alpha _1,\alpha _1 & \alpha _0,\alpha _0 \\ 
& k_1 & k_2 & k_3
\end{array} \right]  \nonumber \\
\lo= \frac{[1+(-1)^{p_1}]2^{2\alpha _0-2}(\alpha _1-\alpha _0)!
\Gamma(\alpha _1+1/2)\prod_{a=1}^{3}\Gamma (p_a/2+\alpha _0
+1/2)}{\Gamma (1/2)\Gamma \left( (k_1+k_2+k_3)/2+\alpha _1
+3/2\right) }  \nonumber \\
\times \frac{\Gamma (\alpha _1+1+k_1)\Gamma (\alpha _1+1+k_2)
\Gamma (\alpha _0+1+k_3)}{\Gamma (2\alpha _1+1+k_1)\Gamma (2
\alpha _1+1+k_2)\Gamma (2\alpha _0+1+k_3)}  \nonumber \\
\times \sum_{u}\frac{\left( (k_1+k_2+k_3)/2+2\alpha _1
-u\right) !}{u!(\alpha _1-\alpha _0-u)!(p_1/2-u)!(p_2/2-u)!
(p_3/2+\alpha _0-\alpha _1+u)!}  \nonumber \\
\times \frac{(-1)^{u}}{\Gamma (\alpha _0+1/2+u)\Gamma (\alpha _1+1/2-u)}  
\label{iJpsr}
\eea
in terms of the balanced (Saalsch\"{u}tzian) $_4F_3(1)$ type series 
\cite{Sl66,GR90}. Here 
\[
p_i=k_j+k_{k}-k_i+2p_i^{\prime },\qquad p_1^{\prime }=
p_2^{\prime }=0,\qquad p_3^{\prime }=\alpha _1-\alpha _0 
\]
are integers.

\section{Canonical basis states and coupling coefficients of SO($n$)}

The canonical basis states of the symmetric (class-one) irreducible
representation $l=l_{(n)}$ for the chain SO$(n)\!\supset $SO$(n-1)\supset
\cdot \cdot \cdot \supset $SO(3)$\supset $SO(2) are labelled by the 
$(n-2)$-tuple $M=(l_{(n-1)},N)=(l_{(n-1)},...,l_{(3)},m_{(2)})$ of 
integers 
\begin{equation}
l_{(n)}\geq l_{(n-1)}\geq ...\geq l_{(3)}\geq |m_{(2)}|.  \label{cnl}
\end{equation}
The dimension of representation space is 
\begin{equation}
d_l^{(n)}=\frac{(2l+n-2)(l+n-3)!}{l!(n-2)!}.  \label{dimo}
\end{equation}

Special matrix elements $D_{M0}^{n,l}(g)$ of SO($n$) irreducible
representation $l_{(n)}=l$ with zero for the $(n-2)$-tuple (0,...,0) 
depend only on the rotation (Euler) angles $\theta _{n-1},\theta _{n-2},%
...,\theta _2,\theta _1$ (coordinates on the unit sphere $S_{n-1}$) 
and may be factorized as 
\begin{equation}
D_{M0}^{n,l}(g)=t_{l^{\prime }0}^{n,l}(\theta _{n-1})D_{N0}^{n-1,
l^{\prime }}(g^{\prime }).  \label{dmel}
\end{equation}
Here $D_{N0}^{n-1,l^{\prime }}(g^{\prime })$ are the matrix elements of 
SO($n-1$) irrep $l_{(n-1)}=l^{\prime }$ (with coordinates on the unit 
sphere $S_{n-2}$). Special matrix elements of SO($n$) ($n>3$) irreducible
representation $l_{(n)}=l$ with the SO($n-1$) irrep labels 
$l_{(n-1)}=l^{\prime }$ and 0 and SO($n-2$) label $l_{(n-2)}=0$ for 
rotation with angle $\theta _{n-1}$ in the $(x_n,x_{n-1})$ plane are 
written in terms of the Gegenbauer polynomials as follows: 
\bea
\fl t_{l^{\prime }0}^{n,l}(\theta _{n-1})=\left[ \frac{l!(l-l^{\prime
})!(n-3)!(l^{\prime }+n-4)!(2l^{\prime }+n-3)}{l^{\prime }!(l+l^{\prime
}+n-3)!(l+n-3)!}\right] ^{1/2}  \nonumber \\
\times (n/2-1)_{l^{\prime }}2^{l^{\prime }}\sin ^{l^{\prime }}\theta
_{n-1}C_{l-l^{\prime }}^{l^{\prime }+n/2-1}(\cos \theta _{n-1}),
\label{tmelc}
\eea
(see \cite{Vi65a}) and corresponds to the wavefunction 
$\Psi _{k,l^{\prime }}^{c}(\theta )=\Psi _{l-l^{\prime },l^{\prime }}^{l
+(n-3)/2}(\theta )$ of the tree technique (of the type 2b, see (2.4) of 
\cite{IPSW01}) with factor 
\[
\left[ \frac{\Gamma ((n-1)/2)\sqrt{\pi }\,d_{l^{\prime }}^{(n-1)}}{\Gamma
(n/2)\,d_l^{(n)}}\right] ^{1/2}, 
\]
for appropriate normalization in the case of integration over the group
volume ($0\leq \theta \leq \pi $) with measure ${\rm B}^{-1}\left(
(n-1)/2,1/2\right) \sin ^{n-2}\theta \d \theta $. The remaining 
Euler angles are equal to 0 for the matrix element (\ref{tmelc}). In the 
case of SO(3), we obtain 
\begin{equation}
D_{m0}^{3,l}(\theta _2,\theta _1)=(-1)^{(l^{\prime }-m)/2}t_{l^{\prime }0
}^{3,l}(\theta _2){\rm e}^{\i m\theta _1},\;l^{\prime }=|m|,
\label{tmelc3}
\end{equation}
in accordance with the relation \cite{Vi65a} between the associated 
Legendre polynomials $P_l^{m}(x)$ and special Gegenbauer polynomials 
$C_{l-m}^{m+1/2}(x)$ and the behavior of $P_l^{m}(x)$ under the 
reflection of $m$.

The corresponding $3j$-symbols for the chain SO$(n)\!\supset $SO$(n-1)
\supset \cdot \cdot \cdot \supset $SO(3)$\supset $SO(2) (denoted by 
brackets with simple subscript $n$ and labelled by sets 
$M_a=(l_a^{\prime },N_a)$) may be factorized as follows: 
\numparts \bea 
\fl \left( \begin{array}{ccc}
l_1 & l_2 & l_3 \\ 
M_1 & M_2 & M_3
\end{array} \right) _{\!n}  \nonumber \\
\lo= \left( \begin{array}{ccc}
l_1 & l_2 & l_3 \\ 0 & 0 & 0
\end{array}
\right) _{\!n}^{-1}\int\limits_{{\rm SO}(n)}\d %
gD_{M_10}^{n,l_1}(g)D_{M_20}^{n,l_2}(g)D_{M_30}^{n,l_3}(g)
\label{c3jci} \\
\lo= \left( \begin{array}{ccc}
l_1 & l_2 & l_3 \\ 
l_1^{\prime } & l_2^{\prime } & l_3^{\prime }
\end{array} \right) _{\!(n:n-1)}\left( \begin{array}{ccc}
l_1^{\prime } & l_2^{\prime } & l_3^{\prime } \\ 
N_1 & N_2 & N_3
\end{array} \right) _{\!n-1}.  \label{c3jcf}
\eea \endnumparts
Here the isoscalar factors of $3j$-symbol for the restriction SO$(n)
\supset $SO($n-1)$ are denoted by brackets with composite subscript 
$(n:n-1)$ and are expressed in terms of integrals (\ref{iGpa}) or 
(\ref{iGpc}) involving triplets of the Gegenbauer polynomials, 
\bea
\fl \left( \begin{array}{ccc}
l_1 & l_2 & l_3 \\ 
l_1^{\prime } & l_2^{\prime } & l_3^{\prime }
\end{array} \right) _{\!(n:n-1)}=\left( \begin{array}{ccc}
l_1 & l_2 & l_3 \\ 0 & 0 & 0
\end{array} \right) _{\!n}^{-1}\left( \begin{array}{ccc}
l_1^{\prime } & l_2^{\prime } & l_3^{\prime } \\ 
0 & 0 & 0
\end{array} \right) _{\!n-1}  \nonumber \\
\times \left[ \frac{\Gamma \left( (n-1)/2\right) }{\pi ^{5/2}\Gamma 
(n/2)}\right] ^{1/2}\prod_{a=1}^{3}{\cal N}_{l_a;l_a^{\prime },
\delta _a}^{(n:n-1)}\!\left[ \frac{d_{l_a^{\prime }}^{(n-1)}}{d_{l_a}^{%
(n)}}\right] ^{1/2}  \nonumber \\
\times \int\limits_0^{\pi }\d \theta (\sin \theta )^{l_1^{\prime }
+l_2^{\prime }+l_3^{\prime }+n-2}\prod_{i=1}^{3}C_{l_i-l_i^{\prime 
}}^{l_i^{\prime }+n/2-1}(\cos \theta ),  \label{isfc}
\eea
where 
\begin{equation}
{\cal N}_{l_a;l_a^{\prime },\delta _a}^{(n:n-1)}=
2^{l_a^{\prime }+n/2-2}\Gamma (l_a^{\prime }+n/2-1) 
\left[ \frac{(l_a-l_a^{\prime })!(2l_a+n-2)}{(l_a+l_a^{\prime }
+n-3)!}\right] ^{1/2}  \label{ncc}
\end{equation}
are normalization factors and particular $3j$-symbols 
\bea
\fl \left( \begin{array}{ccc}
l_1 & l_2 & l_3 \\ 0 & 0 & 0
\end{array} \right) _{\!n} =(-1)^{\psi _n}\frac{1}{\Gamma (n/2)}  
\left[ \frac{(J+n-3)!}{(n-3)!\Gamma (J+n/2)}\right.  \nonumber \\
\times \left. \prod_{i=1}^{3}\frac{\left( l_i+n/2-1\right) \Gamma (J-l_i
+n/2-1)}{d_{l_i}^{(n)}(J-l_i)!}\right] ^{1/2}  \label{isf0}
\eea
(vanishing for $J=\frac 12(l_1+l_2+l_3)$ half-integer) are derived in 
\cite{Ju93} (see also special Clebsch--Gordan coefficients 
\cite{KK73,NAl74a}). Equation (\ref{isfc}) together with (\ref{iGpa}) is 
equivalent to the result of \cite{Ju93}, but its most convenient form is 
obtained when the special integral is expressed by means of double-sum 
expression (\ref{iGpc}) (for $i=1,2$ or 3, minimizing (\ref{ntc})), 
which ensure its finite rational structure for fixed shift 
$\frac 12(l_1+l_2-l_3)$ of parameters. In the case of $l_i^{\prime }=0$, 
expression (\ref{iGpr}) for special integral is more convenient, in 
accordance with \cite{Al87}. In (\ref{isf0}), $J-l_i$ ($i=1,2,3$) and $J$ 
are non-negative integers and $\psi _3=J$, in accordance with the angular 
momentum theory \cite{JB77,VMK88}. We may take also 
\begin{equation}
\psi _n=J  \label{psn}
\end{equation}
(see \cite{Ju93}) for $n\geq 4$, in order to obtain the isofactors 
(\ref{isfc}) positive where the maximal values of parameters 
$l_1^{\prime }=l_1,l_2^{\prime }=l_2,l_3^{\prime }=l_3$.

Only by taking into account the phase factor $(-1)^{(l^{\prime }-m)/2}$ of 
(\ref{tmelc3}), we can obtain the consistent signs of the usual Wigner 
coefficients ($3j$-symbols) 
\begin{equation}
\left( \begin{array}{ccc}
l_1 & l_2 & l_3 \\ 
l_1^{\prime } & l_2^{\prime } & l_3^{\prime }
\end{array} \right) _{\!(3:2)}=\left( \begin{array}{ccc}
l_1 & l_2 & l_3 \\ 
m_1 & m_2 & m_3
\end{array} \right) ,  \label{w3jc}
\end{equation}
of SO(3) or SU(2) (where $l_a^{\prime }=|m_a|$ and $m_1+m_2+m_3=0$), with 
\begin{equation}
\left( \begin{array}{ccc}
l_1^{\prime } & l_2^{\prime } & l_3^{\prime } \\ 
0 & 0 & 0
\end{array}
\right) _{\!2}=\delta _{\max (l_1^{\prime },l_2^{\prime },l_3^{\prime 
}),(l_1^{\prime }+l_2^{\prime }+l_3^{\prime })/2}(-1)^{%
(l_1^{\prime }+l_2^{\prime }+l_3^{\prime })/2}  \label{isf0o}
\end{equation}
consequently appearing in (\ref{isfc}) for $n=3$.\footnote{Of course, 
in this case the usual expressions \cite{Vi65a,JB77,VMK88} of the 
Clebsch--Gordan or Wigner coefficients of SU(2) are more preferable in 
comparison with equation (\ref{isfc}).}

We may write (cf \cite{Ju93}) the following dependence between special
Clebsch--Gordan coefficients (denoted by square brackets with subscript) 
and $3j$-symbols of SO($n$): 
\numparts \bea
\fl \left[ \begin{array}{ccc}
l_1 & l_2 & l_3 \\ 
M_1 & M_2 & M_3
\end{array} \right] _{\!n}\left[ \begin{array}{ccc}
l_1 & l_2 & l_3 \\ 
0 & 0 & 0
\end{array} \right] _{\!n}\equiv \langle l_1M_1;l_2M_2|(l_1l_2)l_3M_3
\rangle _n\langle (l_1l_2)l_3 0|l_1 0;l_2 0\rangle _n  \nonumber \\
\lo= d_l^{(n)}\int\limits_{{\rm SO}(n)}\d gD_{M_10}^{n,
l_1}(g)D_{M_20}^{n,l_2}(g)\overline{D_{M_30}^{n,l_3}(g)}  
\label{cgci} \\
\lo= d_l^{(n)}(-1)^{l_3-m_3}\left( \begin{array}{ccc}
l_1 & l_2 & l_3 \\ 
M_1 & M_2 & \overline{M_3}
\end{array} \right) _{\!n}\left( \begin{array}{ccc}
l_1 & l_2 & l_3 \\ 0 & 0 & 0
\end{array} \right) _{\!n},  \label{cgcw}
\eea \endnumparts
where the $(n-2)$-tuple $\overline{M_3}$ is obtained from the 
$(n-2)$-tuple $M_3$ after reflection of the last parameter $m_3$. 
Then in the phase system with $\psi _n=J$, we obtain the following 
relation for the isofactors of CG coefficients in the canonical basis:
\begin{equation}
\fl \left[ \begin{array}{ccc}
l_1 & l_2 & l_3 \\ 
l_1^{\prime } & l_2^{\prime } & l_3^{\prime }
\end{array} \right] _{\!(n:n-1)}=(-1)^{l_3-l_3^{\prime }}\left[ 
\frac{d_{l_3}^{(n)}}{d_{l_3^{\prime }}^{(n-1)}}\right] ^{1/2}\left( 
\begin{array}{ccc}
l_1 & l_2 & l_3 \\ 
l_1^{\prime } & l_2^{\prime } & l_3^{\prime }
\end{array} \right) _{\!(n:n-1)},  \label{isfcg}
\end{equation}
which, together with (\ref{isfc}), (\ref{isf0}), (\ref{psn}) and 
(\ref{iGpc}) or (\ref{iGpb}) substituted by (\ref{iJpb}), allows us 
to obtain expressions for isofactors of SO$(n)\!\supset $SO($n-1$) 
derived in \cite{NAl74a} and satisfying the same phase conditions.

However, as it was noted in \cite{NAl74b}, the choice (\ref{psn}) of 
$\psi _n$ does not give the correct phases for special isofactors of 
SO(4) \cite{B61} in terms of $9j$ coefficients of SU(2) \cite{JB77} 
and for isofactors of SO(5)$\supset $SO(4), as considered in 
\cite{AlJ71,AlJ69,H65}. The contrast of the phases is caused by the 
fact that the signs of the matrix elements of infinitesimal operators 
\[
A_{k,k-1}=x_k\frac{\partial }{\partial x_{k-1}}-x_{k-1}
\frac{\partial }{\partial x_k},\qquad k=3,...,n, 
\]
(with exception of $A_{2,1}$) between the basis states \cite{Vi65a} of 
SO($n$) in terms of Gegenbauer polynomials (in $x_k/r_k$, 
$r_k^2=x_1^2+...+x_k^2$ variables) are opposite to the signs 
of the standard (Gel'fand--Tsetlin) matrix elements [34--36]. 
%\cite{GC50b,BR77,Kl79}.
We eliminate this difference of phases and our results match with the 
isofactors for decomposition of the general and vector irreps 
$m_n \otimes 1$ of SO($n$) \cite{Kl79,GKl77} (specified also in 
\cite{PC88,SR98}) after we multiply isofactors of CG coefficients for the 
restriction SO$(n)\!\supset $SO($n-1$) ($n\geq 4$), i.e.\ the left-hand 
side of (\ref{isfcg}), by 
\[
(-1)^{(l_1+l_2-l_3-l_1^{\prime }-l_2^{\prime }
+l_3^{\prime })/2} 
\]
(cf \cite{NAl74b}), i.e.\ after we omit the phase factors $(-1)^{\psi
_{n}}$ and $(-1)^{\psi _{n-1}}$ in the both auxiliary $3j$-symbols of 
(\ref{isfc}) and $(-1)^{l_3-l_3^{\prime }}$ in relation 
(\ref{isfcg}), again keeping the isofactors (\ref{isfcg}) with the 
maximal values of parameters $l_1^{\prime }=l_1,l_2^{\prime }=
l_2,l_3^{\prime }=l_3$ for this restriction positive. In the both 
phase systems of the factorized SO($n$) CG coefficients ($3j$-symbols) 
the last factors coincide with the usual CG coefficients ($3j$-symbols) 
of angular momentum theory \cite{JB77,VMK88}.

\section{Semicanonical bases and coupling coefficients of SO($n$)}

Furthermore, going to the semicanonical basis of the symmetric 
(class-one) irreducible representation $l$ for the chain 
SO$(n)\!\supset $SO$(n^{\prime })\times $SO$(n^{\prime \prime %
})\!\supset $SO$(n^{\prime }-1)\times $SO$(n^{\prime \prime }-%
1)\!\supset \cdot \cdot \cdot $, respectively, we introduce special 
matrix elements $D_{l^{\prime }M^{\prime },l^{\prime \prime }
M^{\prime \prime };0}^{n:n^{\prime },n^{\prime \prime };l}(g)$ depending 
only on the rotation angles $\theta _{n^{\prime }-1}^{\prime },...%
\theta _1^{\prime }$ and $\theta _{n^{\prime \prime }-1}^{\prime \prime %
},...\theta _1^{\prime \prime }$ of subgroups SO($n^{\prime }$) and 
SO($n^{\prime \prime }$) and the rotation angle $\theta _c$ in 
$(x_n,x_{n^{\prime }})$ plane, with the second matrix index taken to be 
zero as the $(n-2)$-tuple (0,...,0) for scalar of SO($n-1$). These matrix 
elements may be factorized as follows: 
\begin{equation}
D_{l^{\prime }M^{\prime },l^{\prime \prime }M^{\prime \prime
};0}^{n:n^{\prime },n^{\prime \prime };l}(g)=t_{(n^{\prime })l^{\prime
}0,(n^{\prime \prime })l^{\prime \prime }0;\,(n-1)0}^{(n)\,l}(\theta
_c)D_{M^{\prime }0}^{n^{\prime },l^{\prime }}(g^{\prime })D_{M^{\prime
\prime }0}^{n^{\prime \prime },l^{\prime \prime }}(g^{\prime \prime }).
\label{dmeln}
\end{equation}

Instead of the wavefunction $\Psi _{k,l^{\prime \prime },l^{\prime
}}^{b,a}(\theta _c)=\Psi _{(l-l^{\prime }-l^{\prime \prime })/2,
l^{\prime \prime },l^{\prime }}^{l^{\prime \prime }+n^{\prime \prime }/2
-1,l^{\prime }+n^{\prime }/2-1}(\theta _c)$ (of the type 2c, see (2.6) 
of \cite{IPSW01}) of the tree technique after renormalization with factor 
\[
\left[ \frac{\Gamma (n^{\prime }/2)\Gamma (n^{\prime \prime
}/2)\,d_{l^{\prime }}^{(n^{\prime })}d_{l^{\prime \prime }}^{(n^{\prime
\prime })}}{2\,\Gamma (n/2)\,d_l^{(n)}}\right] ^{1/2} 
\]
for the integration over the group volume ($0\leq \theta _c\leq \pi /2$)
with measure 
\[
2{\rm B}^{-1}(n^{\prime }/2,n^{\prime \prime }/2)\sin ^{n^{\prime 
\prime }-1}\theta _c\cos ^{n^{\prime }-1}\theta _c\d 
\theta _c, 
\]
we obtain special matrix elements of the SO($n$) irreducible 
representation $l$ in terms of the Jacobi polynomials 
\bea
\fl t_{(n^{\prime })l^{\prime }0,(n^{\prime \prime })l^{\prime \prime
}0;\,(n-1)0}^{(n)\,l}(\theta _c)=(-1)^{\varphi _{n^{\prime }
n^{\prime \prime }}}\left[ \frac{d_{l^{\prime }}^{(n^{\prime })}
d_{l^{\prime \prime }}^{(n^{\prime \prime })}\Gamma (n/2)}{d_l^{(n)}
\Gamma (n^{\prime }/2)\Gamma (n^{\prime \prime }/2)\,}\right] ^{1/2}
{\cal N}_{l:l^{\prime },l^{\prime \prime }}^{(n:n^{\prime },
n^{\prime \prime })}  \nonumber \\
\times \sin ^{l^{\prime \prime }}\theta _c\cos ^{l^{\prime }}
\theta _cP_{(l-l^{\prime }-l^{\prime \prime })/2}^{(l^{\prime \prime }
+n^{\prime \prime }/2-1,l^{\prime }+n^{\prime }/2-1)}(\cos 2\theta _c),  
\label{tmeln}
\eea
where the left-hand SO$(n^{\prime })\times $SO$(n^{\prime \prime }$) 
labels are $l^{\prime },l^{\prime \prime }$ ($n^{\prime }
+n^{\prime \prime }=n$), the left-hand SO$(n^{\prime }-1)\times $SO%
$(n^{\prime \prime}-1)$ and right-hand SO($n-1$) labels are 0 for 
rotation with angle $\theta _c$ in $(x_n,x_{n^{\prime }})$ plane. Here 
phase $\varphi _{n^{\prime }n^{\prime \prime }}=0$, unless 
$n^{\prime \prime }=2$, or $n^{\prime }=2$, when the left-hand side 
should be replaced, respectively, by $t_{(n-2)l^{\prime }0,(2)
m^{\prime \prime };\;(n-1)0}^{(n)\;l}(\theta _c)$ with 
$l^{\prime \prime }=|m^{\prime \prime }|$, or by 
$t_{(2)m^{\prime },(n-2)l^{\prime \prime }0;\;(n-1)0}^{(n)\;l}(
\theta _c)$ with $l^{\prime }=|m^{\prime }|$ and
\[
\varphi _{n^{\prime }n^{\prime \prime }}=\case 12[\delta _{n^{\prime
\prime }2}(l^{\prime \prime }-m^{\prime \prime })+\delta _{n^{\prime
}2}(l^{\prime }-m^{\prime })] 
\]
on the right-hand side and normalization factor 
\begin{equation}
\fl {\cal N}_{l:l^{\prime },l^{\prime \prime }}^{(n:n^{\prime },n^{\prime
\prime })}=\left[ \frac{(l+n/2-1)\left( (l-l^{\prime }-l^{\prime \prime
})/2\right) !\Gamma \left( (l+l^{\prime }+l^{\prime \prime }+n
-2)/2\right) }{\Gamma \left( (l-l^{\prime }+l^{\prime \prime }
+n^{\prime \prime })/2\right) \Gamma \left( (l+l^{\prime }
-l^{\prime \prime }+n^{\prime })/2\right) }\right] ^{1/2}.  \label{ncn}
\end{equation}

The $3j$-symbols for the chain SO$(n)\!\supset $SO$(n^{\prime })\times $%
SO$(n^{\prime \prime })\!\supset $SO$(n^{\prime }-1)\times $SO$(n^{\prime
\prime }-1)\!\supset \cdot \cdot \cdot $, labelled by the sets 
$M_i=(l_i^{\prime },N_i^{\prime };l_i^{\prime \prime },
N_i^{\prime \prime })$ may be factorized as follows: 
\bea
\fl \left( \begin{array}{ccc}
l_1 & l_2 & l_3 \\ 
M_1 & M_2 & M_3
\end{array} \right) _{\!n}=\left( \begin{array}{ccc}
l_1 & l_2 & l_3 \\ 
l_1^{\prime },l_1^{\prime \prime } & l_2^{\prime },l_2^{\prime
\prime } & l_3^{\prime },l_3^{\prime \prime }
\end{array} \right) _{\!(n:n^{\prime }n^{\prime \prime })}  \nonumber \\
\times \left( \begin{array}{ccc}
l_1^{\prime } & l_2^{\prime } & l_3^{\prime } \\ 
N_1^{\prime } & N_2^{\prime } & N_3^{\prime }
\end{array} \right) _{\!n^{\prime }}\left( \begin{array}{ccc}
l_1^{\prime \prime } & l_2^{\prime \prime } & 
l_3^{\prime \prime } \\ N_1^{\prime \prime } & N_2^{\prime \prime 
} & N_3^{\prime \prime } 
\end{array} \right) _{\!n^{\prime \prime }}.  \label{c3jnc}
\eea
Now the SO$(n)\!\supset $SO$(n^{\prime })\times $SO$(n^{\prime \prime })$
isofactor of $3j$-symbol is expressed as follows: 
\bea
\fl \left( \begin{array}{ccc}
l_1 & l_2 & l_3 \\ 
l_1^{\prime },l_1^{\prime \prime } & l_2^{\prime },l_2^{\prime
\prime } & l_3^{\prime },l_3^{\prime \prime }
\end{array} \right) _{\!(n:n^{\prime }n^{\prime \prime })}=\left( 
\begin{array}{ccc}
l_1 & l_2 & l_3 \\ 0 & 0 & 0
\end{array} \right) _{\!n}^{-1}\left( \begin{array}{ccc}
l_1^{\prime } & l_2^{\prime } & l_3^{\prime } \\ 0 & 0 & 0
\end{array} \right) _{\!n^{\prime }}  \nonumber \\
\times \left( \begin{array}{ccc}
l_1^{\prime \prime } & l_2^{\prime \prime } & 
l_3^{\prime \prime } \\ 0 & 0 & 0
\end{array}
\right) _{\!n^{\prime \prime }}\prod_{a=1}^{3}{\cal N}_{l_a;l_a^{\prime },
l_a^{\prime \prime }}^{(n:n^{\prime },n^{\prime \prime })}\left[ %
\frac{d_{l_a^{\prime }}^{(n^{\prime })}d_{l_a^{\prime \prime }}^{%
(n^{\prime \prime })}}{d_{l_a}^{(n)}}\right] ^{1/2}  \nonumber \\
\times {\rm B}^{1/2}(n^{\prime }/2,n^{\prime \prime }/2)\;
\widetilde{\cal I}\left[ \begin{array}{cccc}
\alpha _0,\beta _0 & \alpha _1,\beta _1 & 
\alpha _2,\beta _2 & \alpha _3,\beta _3 \\ 
& k_1 & k_2 & k_3
\end{array}
\right] ,  \label{isfnc}
\eea
in terms of auxiliary $3j$-symbols (\ref{isf0}) of the canonical bases
[turning into phase factors of the type \ref{isf0o} for $n^{\prime }=2$ 
or $n^{\prime \prime }=2$], normalization factors (\ref{ncn}) and the 
integrals involving triplets of Jacobi polynomials 
(\ref{iJp})--(\ref{iJpd}), with parameters 
\begin{eqnarray*}
k_i=\case 12(l_i-l_i^{\prime }-l_i^{\prime \prime }),\qquad 
\alpha _i=l_i^{\prime \prime }+n^{\prime \prime }/2-1,\qquad 
\beta _i=l_i^{\prime }+n^{\prime }/2-1, \\
\alpha _0=n^{\prime \prime }/2-1,\qquad \beta _0=n^{\prime }/2-1
\end{eqnarray*}
and 
\begin{eqnarray*}
p_i^{\prime }=\case 12(l_j^{\prime }+l_k^{\prime }
-l_i^{\prime }),\qquad p_i^{\prime \prime }=\case 12(l_j^{\prime 
\prime }+l_k^{\prime \prime }-l_i^{\prime \prime }), \\ 
p_i=\case 12(l_j+l_k-l_i)\qquad (i,j,k=1,2,3).
\end{eqnarray*}
The number of terms in expansion (\ref{iJpd}) of integrals involving 
triplets of Jacobi polynomials never exceeds 
\numparts \begin{equation}
\fl B_i=\min \left( \case{1}{6}(p_i+1)_3,(p_i^{\prime \prime }+1)(k_j+1)
(k_{k}+1),\case 12(p_i^{\prime \prime }+1)(p_i+1)_2)\right)   
\label{ntnc}
\end{equation}
and decreases in the intermediate region (e.g., when 
$p_i^{\prime \prime }<p_i+1$), described by the volume of the obliquely 
truncated rectangular parallelepiped of 
$(p_i^{\prime \prime }+1)\times (k_j+1)\times (k_{k}+1)$ size. 

In particular, in the case of $n^{\prime \prime }=2$ parameters 
$l_1^{\prime \prime },l_2^{\prime \prime },l_3^{\prime \prime }$ in 
$3j$-symbol (\ref{isfnc}) should be replaced by $m_1^{\prime \prime }
=\pm l_1^{\prime \prime },m_2^{\prime \prime }=\pm l_2^{\prime 
\prime },m_3^{\prime \prime }=\pm l_3^{\prime \prime }$ so that 
$m_1^{\prime \prime }+m_2^{\prime \prime }+m_3^{\prime \prime }=0$.
Hence at least one parameter $p_{i^{\prime }}^{\prime \prime }=0$ and the 
number of terms in the $i^{\prime }$th double sum version of 
(\ref{iJpd}) [related to (\ref{iJra}) and to the Kamp\'{e} de 
F\'{e}riet \cite{K-F21} function $F_{2:1}^{2:2}$] does not exceed 
\begin{equation}
\widetilde{B}_{i^{\prime }}=\min \left( \case 12(p_{i^{\prime
}}+1)_2,(k_{j^{\prime }}+1)(k_{k^{\prime }}+1)\right) ,  \label{ntsc}
\end{equation} \endnumparts
although for small values of $p_i$ so that $B_i<\widetilde{B}%
_{i^{\prime }}$ the $i$th version of (\ref{iJpd}) or (\ref{iJrb}) may 
be more preferable.

We may also express the isofactors of the CG coefficients for restriction 
SO$(n)\!\supset $SO$(n^{\prime })\times $SO$(n^{\prime \prime })$ in 
terms of the isofactors of $3j$-symbols, 
\begin{equation}
\fl \left[ \begin{array}{ccc}
l_1 & l_2 & l_3 \\ 
\!l_1^{\prime },l_1^{\prime \prime }\! & \!l_2^{\prime },l_2^{\prime
\prime }\! & \!l_3^{\prime },l_3^{\prime \prime }\!
\end{array}
\right] _{(n:n^{\prime }n^{\prime \prime })}=(-1)^{\varphi }\!\left[ %
\frac{d_{l_3}^{(n)}}{\!d_{l_3^{\prime }}^{(n^{\prime })}d_{l_3^{%
\prime \prime }}^{(n^{\prime \prime })\!}}\right] ^{1/2}\!\!\!\left( 
\begin{array}{ccc}
l_1 & l_2 & l_3 \\ 
\!\!l_1^{\prime },l_1^{\prime \prime }\! & \!l_2^{\prime },l_2^{\prime
\prime }\! & \!l_3^{\prime },l_3^{\prime \prime }\!\!
\end{array} 
\right) _{\!\!(n:n^{\prime }n^{\prime \prime })},  \label{isfncg}
\end{equation}
with the phase $\varphi =0$ (since $l_3-l_3^{\prime }-l_3^{\prime
\prime }$ is even), when $\psi _n,\psi _{n^{\prime }},\psi _{n^{\prime
\prime }}$ are taken to be equal to $J,J^{\prime },J^{\prime \prime }$,
respectively, in all the auxiliary $3j$-symbols (\ref{isf0}), in contrast
to 
\[
\varphi =m_3^{\prime \prime }\delta _{n^{\prime \prime }2}+m_3^{\prime
}\delta _{n^{\prime }2}+l_3^{\prime \prime }\delta _{n^{\prime \prime
}3}+l_3^{\prime }\delta _{n^{\prime }3}, 
\]
appearing when $\psi _n$ is taken to be zero for $n\geq 4$. Again we 
need to replace, respectively, for $n^{\prime \prime }=2$ parameters 
$l_1^{\prime \prime },l_2^{\prime \prime },l_3^{\prime \prime }$ 
on the left-hand side by $m_1^{\prime \prime },m_2^{\prime \prime },
m_3^{\prime \prime }$ so that $l_1^{\prime \prime }=|m_1^{\prime 
\prime }|,l_2^{\prime \prime }=|m_2^{\prime \prime }|,
l_3^{\prime \prime }=|m_3^{\prime \prime }|$ (with 
$m_1^{\prime \prime }+m_2^{\prime \prime }=m_3^{\prime \prime }$) 
and in the right-hand side by $m_1^{\prime \prime },m_2^{\prime \prime
},-m_3^{\prime \prime }$, as well as for $n^{\prime }=2$ parameters 
$l_1^{\prime },l_2^{\prime },l_3^{\prime }$ on the left-hand side by 
$m_1^{\prime },m_2^{\prime },m_3^{\prime }$ so that $l_1^{\prime
}=|m_1^{\prime }|,l_2^{\prime }=|m_2^{\prime }|,l_3^{\prime
}=|m_3^{\prime }|$ ($m_1^{\prime }+m_2^{\prime }=m_3^{\prime }$) 
and on the right-hand side by 
$m_1^{\prime },m_2^{\prime },-m_3^{\prime }$.

Regarding the different triple-sum versions (\ref{iGpa})--(\ref{iGpd}) 
of integrals involving triplets of Gegenbauer and Jacobi polynomials and 
comparing expressions (\ref{isfnc}) and (\ref{isfc}), we derive the 
following duplication relation between the generic SO$(n)\!\supset $SO$%
(n-1)$ and special SO$(2n+2)\!\supset $SO$(n-1)\times $SO$(n-1)$ 
isofactors of the $3j$-symbols: 
\bea
\fl \left( \begin{array}{ccc}
2l_1 & 2l_2 & 2l_3 \\ 
\!l_1^{\prime },l_1^{\prime } & \!l_2^{\prime },l_2^{\prime }\! & 
l_3^{\prime },l_3^{\prime }\!
\end{array} \right) _{\!\!(2n-2:n-1,n-1)}=\left( \begin{array}{ccc}
\!2l_1 & \!2l_2\! & 2l_3\! \\ 0 & 0 & 0
\end{array} \right) _{\!\!2n-2}^{-1}\left( \begin{array}{ccc}
l_1 & l_2 & l_3 \\ 
l_1^{\prime } & l_2^{\prime } & l_3^{\prime }
\end{array} \right) _{\!\!(n:n-1)}  \nonumber \\
\times \left( \begin{array}{ccc}
l_1 & l_2 & l_3 \\ 0 & 0 & 0
\end{array} \right) _{\!n}\left( \begin{array}{ccc}
l_1^{\prime } & l_2^{\prime } & l_3^{\prime } \\ 
0 & 0 & 0
\end{array}
\right) _{\!n-1}\prod_{a=1}^{3}\left[ \frac{d_{l_a}^{(n)}d_{l_a^{\prime
}}^{(n-1)}}{d_{2l_a}^{(2n-2)}}\right] ^{1/2},  \label{isfd}
\eea
with auxiliary $3j$-symbols (\ref{isf0}) of the canonical bases and the 
irrep dimensions appearing.

\section{Basis states and coupling coefficients of the class-two (mixed
tensor) representations of U($n$)}

Mixed tensor irreducible representations $[p+q,q^{n-2},0]\equiv [p,%
\dot{0},-q]$ of U($n$) containing scalar irrep $[q^{n-1}]\equiv 
[\dot{0}]$ of subgroup U($n-1$) (with repeating zeros denoted by 
$\dot{0}$) are called class-two irreps \cite{K74,NAl75}; their canonical 
basis states for the chain U$(n)\!\supset $U$(n-1)\times $U$(1)\!\supset 
\cdot \cdot \cdot \supset $U$(2)\times $U$(1)\!\supset $U(1) are labelled 
by the set 
\[
\fl Q_{(n)}=(p_{(n-1)},q_{(n-1)};Q_{(n-1)})=(p_{(n-1)},q_{(n-1)};
p_{(n-2)},q_{(n-2)};...,p_{(2)},q_{(2)};p_{(1)}),
\] 
where 
\[
\fl p=p_{(n)}\geq p_{(n-1)}\geq ...\geq p_{(2)}\geq 0\quad {\rm and}\quad 
q=q_{(n)}\geq q_{(n-1)}\geq ...\geq q_{(2)}\geq 0 
\]
are integers, with $p_{(2)}\geq p_{(1)}\geq -q_{(2)}$ in addition, and
parameters 
\[
\fl M_{(1)}=p_{(1)},\; M_{(2)}=p_{(2)}-q_{(2)}-p_{(1)},...,%
\;M_{(r)}=p_{(r)}-q_{(r)}-p_{(r-1)}+q_{(r-1)} 
\]
which correspond to irreps of subgroups U$(1)$, beginning from the last
one.

The dimension of representation space is 
\begin{equation}
d_{[p,\dot{0},-q]}^{(n)}=\frac{(p+q+n-1)(p+1)_{n-2}(q+1)_{n-2}}{(n-1)!
(n-2)!}.  \label{dimu}
\end{equation}

Special matrix elements $D_{Q_{(n)};0}^{n[p,\dot{0},-q]}(g)$ of U($n$) 
irrep $[p,\dot{0},-q]$ with zero as the second index for the scalar of 
subgroup U($n-1$) depend only on the rotation angles $\varphi _n,
\varphi _{n-1},...,\varphi _2,\varphi _1$, where 0$\leq \varphi _i
\leq 2\pi $ corresponds to the $i$th diagonal subgroup U(1) 
($i=1,2,...,n$), and $\theta _n,\theta _{n-1},...,\theta _3,
\theta _2$, where $0\leq \theta _r\leq \pi /2$, corresponds to the 
transformation 
\[
\left| \begin{array}{cc}
\cos \theta _r & \i \sin \theta _r \\ 
\i \sin \theta _r & \cos \theta _r
\end{array} \right| 
\]
in the plane of $(r-1)$st and $r$th coordinates ($r=2,3,...,n$) and may 
be factorized as follows: 
\begin{equation}
D_{Q_{(n)};0}^{n[p,\dot{0},-q]}(g)={\rm e}^{\i M_n\varphi _n}D_{[p^{\prime 
},\dot{0},-q^{\prime }]0;0}^{n[p,\dot{0},-q]}(\theta _n)D_{Q_{(n-1)};
0}^{n-1[p^{\prime },\dot{0},-q^{\prime }]}(g^{\prime })  \label{dmelu}
\end{equation}
with appropriate normalization in the case of integration over the group
volume with measure 
\[
\frac{(n-1)!}{2\pi ^{n}}\prod_{r=2}^{n}\sin ^{2r-3}\theta _r\cos \theta
_r\d \theta _r\prod_{i=1}^{n}\d \varphi _i. 
\]
Here $D_{Q_{(n-1)};0}^{n-1[p^{\prime },\dot{0},-q^{\prime }]}
(g^{\prime })$ are the matrix elements of U($n-1$) irrep $[p^{\prime },
\dot{0},-q^{\prime }]=[p_{(n-1)},\dot{0},-q_{(n-1)}]$ (with parameters 
obtained after omitting $\varphi _n$ and $\theta _n$). Special matrix 
elements of U($r$) irreducible representation $[p,\dot{0},-q]$ with the 
U($r-1$) irrep labels $[p^{\prime },\dot{0},-q^{\prime }]$ and 0 and 
SU($r-2$) irrep label 0 for rotation with angle $\theta _r$ in the 
$(x_r,x_{r-1})$ plane are written in terms of the $D$-matrices of SU(2) 
as follows: 
\bea
\fl D_{[p^{\prime },\dot{0},-q^{\prime }]0;0}^{r[p,\dot{0},-q]}
(\theta _r)=\left[ (p+q+r-1)\,d_{[p^{\prime },\dot{0},
-q^{\prime }]}^{(r-1)}\right] ^{1/2}\left[ (r-1)\,d_{[p,\dot{0},-q]}^{(r)}
\right] ^{-1/2}  \nonumber \\
\times (\i \sin \theta _r)^{-r+2}P_{p^{\prime }+(q-p+r-2)/2,
-(p-q+r-2)/2-q^{\prime }}^{(p+q+r-2)/2}(\cos 2\theta _r)  \label{tmelud}
\eea
and further, taking into account the identity $P_{m,n}^{l}(x)=
P_{-n,-m}^{l}(x)$, in terms of the Jacobi polynomials 
\bea
\fl D_{[p^{\prime },\dot{0},-q^{\prime }]0;0}^{r[p,\dot{0},-q]}
(\theta _r)={\cal N}_{ [p^{\prime },\dot{0},-q^{\prime }]}^{r[p,\dot{0},
-q]}\left[ d_{ [p^{\prime },\dot{0},-q^{\prime }]}^{(r-1)}
\left( (r-1)\,d_{ [p,\dot{0},-q]}^{(r)}\right) ^{-1}\right] ^{1/2}  
\nonumber \\
\times (\i \sin \theta _r)^{p^{\prime }+q^{\prime }}(\cos 
\theta _r)^{|M|}P_{K}^{(L^{\prime }+r-2,|M|)}(\cos 2\theta _r),  
\label{tmeluj}
\eea
where 
\[
K=\min (p-p^{\prime },q-q^{\prime }),\qquad M=p-q-p^{\prime }+q^{\prime },
\qquad L^{\prime }=p^{\prime }+q^{\prime } 
\]
and 
\numparts \bea
\fl {\cal N}_{ [p^{\prime },\dot{0},-q^{\prime }]}^{r[p,\dot{0},-q]}
=\left[ \frac{(p+q+r-1)K!(p+q+r-2-K)!}{\left( |M|+K\right) !(p+q+r-2-|M|
-K)!}\right] ^{1/2}  \label{ncua} \\
\lo= \left[ \frac{(p+q+r-1)K!(L^{\prime }+r-2+|M|+K)!}{\left( |M|+K%
\right)!(L^{\prime }+r-2+K)!}\right] ^{1/2}.  \label{ncub}
\eea \endnumparts
Factor $\i ^{p^{\prime }+q^{\prime }}$, also appeared in 
\cite{NAl75} (but was absent in the generic expressions of $D$-matrix 
elements \cite{Kl79,Vi74}), ensures the complex conjugation relation 
\begin{equation}
\overline{D_{ [p^{\prime },\dot{0},-q^{\prime }]0;0}^{r[p,\dot{0},-q]}
(\theta _r)}=(-1)^{p^{\prime }+q^{\prime }}D_{ [q^{\prime },\dot{0},
-p^{\prime }]0;0}^{r[q,\dot{0},-p]}(\theta _r),  \label{ccur}
\end{equation}
in accordance with the SU(2) case and the system of phases of Baird and
Biedenharn \cite{BB64}, which is correlated to the positive signs of 
the Gel'fand--Tsetlin matrix elements \cite{BR77,Kl79,GC50a} 
of the U($n$) generators $E_{r,r-1}$. Alternatively, the states 
$\Psi _{p^{\prime }+q^{\prime },p+q,M}(\theta _r)$, as defined in 
\cite{KMSS92,K74} and related to the hyperspherical harmonics, correspond 
to the Jacobi polynomials with interchanged parameters $\alpha $ and 
$\beta $. Hence the variables are mutually reflected [here and in 
\cite{KMSS92,K74} as $\cos 2\theta _r$ and ($-\cos 2\theta _r$)].

Using the integration over group (cf \cite{Kl79,K74,KlG79}), the
corresponding $3j$-symbols of the class-two irreps for the chain 
U$(n)\!\supset $U$(n-1)\times $U$(1)\!\supset \cdot \cdot \cdot \supset $%
U$(2)\times $U$(1)\!\supset $U(1) may be factorized as follows: 
\numparts \bea
\fl \sum_{\rho }\left( \begin{array}{ccc}
\![p_1,\dot{0},-q_1] & [p_2,\dot{0},-q_2] & 
[p_3,\dot{0},-q_3] \\ 
Q_{1(n)} & Q_{2(n)} & Q_{3(n)}
\end{array} \right) _{\!n}^{\rho }  \nonumber \\
\times \left( \begin{array}{ccc}
\![p_1,\dot{0},-q_1] & [p_2,\dot{0},-q_2] & 
[p_3,\dot{0},-q_3] \\ 
\ [\dot{0}] & [\dot{0}] & [\dot{0}]
\end{array}
\right) _{\!n}^{\rho }  \nonumber \\
\lo= \int\limits_{{\rm U}(n)}\d gD_{Q_{1(n)};0}^{n[p_1,
\dot{0},-q_1]}(g)D_{Q_{2(n)};0}^{n[p_2,\dot{0},-q_2]}(g)
D_{Q_{3(n)};0}^{n[p_3,\dot{0},-q_3]}(g)  \label{c3jui} \\
\lo= \delta _{p_1+p_2+p_3,q_1+q_2+q_3}(-1)^{p_1^{\prime
}+p_2^{\prime }+p_3^{\prime }+K_1+K_2+K_3}(n-1)^{-1/2}  
\nonumber \\
\times \prod_{a=1}^{3}{\cal N}_{ [p_a^{\prime },\dot{0},
-q_a^{\prime }]}^{n[p_a,\dot{0},-q_a]}\left[ d_{ [p_a^{\prime },
\dot{0},-q_a^{\prime }]}^{(n-1)}\left( d_{ [p_a,\dot{0},-q_a]}^{(n)}%
\right) ^{-1}\right] ^{1/2}  \nonumber \\
\times \widetilde{\cal I}\left[ \begin{array}{cccc}
0,n\!-\!2 & \!|M_1|,L_1^{\prime }\!+\!n\!-\!2 & \!|M_2|,L_2^{\prime }\!%
+\!n\!-\!2 & \!|M_3|,L_3^{\prime }\!+\!n\!-\!2 \\ 
& K_1 & K_2 & K_3
\end{array} \right]   \nonumber \\
\times \sum_{\rho ^{\prime }}\left( \begin{array}{ccc}
\![p_1^{\prime },\dot{0},-q_1^{\prime }] & [p_2^{\prime },
\dot{0},-q_2^{\prime }] & [p_3^{\prime },\dot{0},-q_3^{\prime }] \\ 
Q_{1(n)}^{\prime } & Q_{2(n)}^{\prime } & Q_{3(n)}^{\prime }
\end{array} \right) _{\!n-1}^{\rho ^{\prime }}  \nonumber \\
\times \left( \begin{array}{ccc}
\![p_1^{\prime },\dot{0},-q_1^{\prime }] & [p_2^{\prime },
\dot{0},-q_2^{\prime }] & [p_3^{\prime },\dot{0},-q_3^{\prime }] \\ 
\ [\dot{0}] & [\dot{0}] & [\dot{0}]
\end{array} \right) _{\!n-1}^{\rho ^{\prime }}.  \label{c3juf}
\eea \endnumparts
Here $\rho $ and $\rho ^{\prime }$ are the multiplicity labels of the 
U$(n)$ and U$(n-1)$ scalars in the decompositions $[p_1,\dot{0},
-q_1]\otimes [p_2,\dot{0},-q_2]\otimes [p_3,\dot{0},-q_3]$ and 
$[p_1^{\prime },\dot{0},-q_1^{\prime }]\otimes [p_2^{\prime },
\dot{0},-q_2^{\prime }]\otimes [p_3^{\prime },\dot{0},
-q_3^{\prime }]$. The integral involving the product of three Jacobi 
polynomials that appeared in (\ref{c3juf}) also corresponds to the 
SO$(2n)\!\supset $SO$(2n-2)\times $SO(2) isofactor of $3j$-symbol 
\[
\left( \begin{array}{ccc}
p_1+q_1 & p_2+q_2 & p_3+q_3 \\ 
p_1^{\prime }+q_1^{\prime },M_1 & p_2^{\prime }+q_2^{\prime },
M_2 & p_3^{\prime }+q_3^{\prime },M_3
\end{array} \right) _{(2n:2n-2,2)},
\]
considered in previous section and may be expressed (after some 
permutation of parameters) as double sum by means of (\ref{iJra}) or 
(\ref{iJrb}). For normalization of the corresponding $3j$-symbols of 
U$(n)\!\supset $U$(n-1)$ we may use square root of 
\bea
\fl \sum_{\rho }\left[ \left( \begin{array}{ccc}
\![p_1,\dot{0},-q_1] & [p_2,\dot{0},-q_2] & 
[p_3,\dot{0},-q_3] \\ 
\ [\dot{0}] & [\dot{0}] & [\dot{0}]
\end{array} \right) _{\!n}^{\rho }\right] ^{2}=\delta _{p_1+p_2+p_3,
q_1+q_2+q_3}  \nonumber \\
\times (-1)^{\min (p_1,q_1)+\min (p_2,q_2)+\min (p_3,q_3)}\frac{(n-1)!
[(n-2)!]^2}{\prod_{a=1}^{3}\left( \min (p_a,q_a)+1\right) _{\!n-2}}  
\nonumber \\
\times \widetilde{\cal I}\left[ \begin{array}{cccc}
0,n\!-\!2 & \!|p_1\!-\!q_1|,n\!-\!2 & \!|p_2\!-\!q_2|,n\!-\!2 & 
\!|p_3\!-\!q_3|,n\!-\!2 \\ 
& \min (p_1,q_1) & \min (p_2,q_2) & \min (p_3,q_3)
\end{array} \right] ,  \label{isfu0}
\eea
with non-vanishing extreme $3j$-symbols in the left-hand side for a 
single value of the multiplicity label $\rho $, which is not correlated 
with the canonical [46--48] %\cite{BLCC72,BLL85,LBL88} 
and other (see [49--51]) %\cite{Al88,Al95,Al96}) 
external labeling schemata of the coupling coefficients of U($n$). In 
contrast to the particular $3j$-symbols (\ref{isf0}) of SO($n$), equation 
(\ref{isfu0}) is summable only in the multiplicity-free cases. In 
addition to three double-sum versions of (\ref{iJra}) and (\ref{iJrb}),
integral on the right-hand side of (\ref{isfu0}) also may be expressed as 
three different double-sum series by means of (\ref{iJpd}), taking into 
account the symmetry relation (\ref{iJpt}). Of course, (\ref{isfu0}) is
always positive as an analogue of denominator function of the SU(3) 
canonical tensor operators [46--48,~52]. %\cite{BLCC72,BLL85,LBL88,Al99}.

Taking into account (\ref{ccur}) we may also obtain expression for the
Clebsch--Gordan coefficients of class-two representation of U($n$) 
\bea
\fl \sum_{\rho }\left[ \begin{array}{cc}
\![p_1,\dot{0},-q_1] & [p_2,\dot{0},-q_2] \\ 
Q_{1(n)} & Q_{2(n)}
\end{array} \left| \begin{array}{c}
\![p,\dot{0},-q] \\ Q_{(n)}
\end{array} \right. \right] _{\!n}^{\rho }  \nonumber \\
\times \left[ \begin{array}{cc}
\![p_1,\dot{0},-q_1] & [p_2,\dot{0},-q_2] \\ 
\ [\dot{0}] & [\dot{0}]
\end{array} \left| \begin{array}{c}
\![p,\dot{0},-q] \\ \![\dot{0}]
\end{array} \right. \right] _{\!n}^{\rho }  \nonumber \\
\lo= d_{ [p,\dot{0},-q]}^{(n)}\int\limits_{{\rm U}(n)}\d %
gD_{Q_{1(n)};0}^{n[p_1,\dot{0},-q_1]}(g)D_{Q_{2(n)};0}^{n[p_2,
\dot{0},-q_2]}(g)\overline{D_{Q_{(n)};0}^{n[p,\dot{0},-q]}(g)},  
\label{cgcuf}
\eea
with the integrals involving the product of three Jacobi polynomials and 
the CG coefficients of U($n-1$) of the same type and some phase and irrep 
dimension factors. Particularly, we obtain the following expression for 
isofactors of special SU(3) Clebsch--Gordan coefficients (which perform 
the coupling of the SU(3)-hyperspherical harmonics): 
\bea
\fl \left[ \begin{array}{ccc}
(a^{\prime }b^{\prime }) & (a^{\prime \prime }b^{\prime \prime }) & 
(ab)_0 \\ 
(z^{\prime })i^{\prime } & (z^{\prime \prime })i^{\prime }; & (z)i
\end{array} \right]  \nonumber \\
\lo= \delta _{a^{\prime }+a^{\prime \prime }-a,b^{\prime }+b^{\prime 
\prime }-b}(-1)^{i^{\prime }+i^{\prime \prime }-i+K_1+K_2-K}\,
\frac 12\left[ \frac{(2i^{\prime }+1)(2i^{\prime \prime }+1)}{(2i+1)\,
d_{(a^{\prime }b^{\prime })}^{(3)}d_{(a^{\prime \prime }b^{\prime \prime 
})}^{(3)}}\right] ^{1/2}  \nonumber \\
\times \left\{ \frac{(-1)^{\min (a^{\prime },b^{\prime })+\min (%
a^{\prime \prime },b^{\prime \prime })+\min (a,b)}}{\left( \min (%
a^{\prime },b^{\prime })+1\right) \left( \min (a^{\prime \prime },
b^{\prime \prime })+1\right) \left( \min (a,b)+1\right) }\right.  
\nonumber \\
\times \left. \widetilde{\cal I}\left[ \begin{array}{cccc}
0,1 & |a^{\prime }-b^{\prime }|,1 & |a^{\prime \prime }
-b^{\prime \prime }|,1 & |a-b|,1 \\ 
& \min (a^{\prime },b^{\prime }) & \min (a^{\prime \prime },
b^{\prime \prime }) & \min (a,b)
\end{array} \right] \right\} ^{-1/2}  \nonumber \\
\times {\cal N}_{ [i^{\prime }-z^{\prime },-i^{\prime }-z^{\prime
}]}^{3[a^{\prime },0,-b^{\prime }]}{\cal N}_{ [i^{\prime \prime
}-z^{\prime \prime },-i^{\prime \prime }-z^{\prime \prime }]}^{3[%
a^{\prime \prime },0,-b^{\prime \prime }]}{\cal N}_{ [i-z,-i-z]}^{3%
[a,0,-b]}\,\left[ \begin{array}{ccc}
i^{\prime } & i^{\prime \prime } & i \\ 
z^{\prime } & z^{\prime \prime } & z
\end{array} \right]  \nonumber \\
\times \widetilde{\cal I}\left[ \begin{array}{cccc}
0,1 & |M^{\prime }|,2i^{\prime }+1 & |M^{\prime \prime }|,2i^{\prime 
\prime }+1 & |M|,2i+1 \\ 
& K^{\prime } & K^{\prime \prime } & K
\end{array} \right] .  \label{isfu3}
\eea
Here $M=a-b+2z,\,K=\min (a+z-i,b-z-i)$ in the notation of 
\cite{Al88,Al96}, with $(a~b)$ for the mixed tensor irreps, where 
$a=p_{(3)},\,b=q_{(3)}$ and the basis states are labelled by the isospin 
$i=\case 12(p_{(2)}+q_{(2)})$, its projection $i_{z}=p_{(1)}-
\frac 12(p_{(2)}-q_{(2)})$ and the parameter 
$z=\case 13(b-a)-\case 12y=\case 12(q_{(2)}-p_{(2)})$ instead of 
the hypercharge $y=p_{(2)}-q_{(2)}-\frac 23(p_{(3)}-q_{(3)})$.

\section{Concluding remarks}

In this paper, we reconsidered the $3j$-symbols and Clebsch--Gordan
coefficients of the orthogonal SO($n$) and unitary U($n$) groups for all
three representations corresponding to the (ultra)spherical or 
hyperspherical harmonics of these groups (i.e.\ irreps induced 
\cite{BR77} by the scalar representations of the SO($n-1$) and U($n-1$) 
subgroups, respectively). For the corresponding isoscalar factors of the 
$3j$-symbols and coupling coefficients, the ordinary integrations 
involving triplets of the Gegenbauer and the Jacobi polynomials yield 
the most symmetric triple-sum expressions, however without the apparent 
triangle conditions. These conditions are visible and efficient only in 
expressions of the type \cite{NAl74a,Al83} derived after complicated 
analytical continuation procedure of special 
Sp(4)$\supset $SU(2)$\times $SU(2) isofactors (cf \cite{AlJ71,AlJ69}). 
Actually, only for a fixed integer shift parameter 
$p_i=\frac 12(l_j+l_k-l_i)$ it is evident that the corresponding 
integrals involving triplets of the Gegenbauer and the Jacobi polynomials 
are rational functions of remaining parameters. Practically, the concept 
of the canonical unit tensor operators (see section 21 of chapter 3 of 
\cite{BL81}) for symmetric irreps of SO($n$) may be formulated only under 
such a condition.

Similarly as special terminating double-hypergeometric series of Kamp\'{e}
de F\'{e}riet-type [19--21,~53] %\cite{Al00,K-F21,LV-J01,V-JPR94} 
correspond to the stretched $9j$ coefficients of SU(2), the definite 
terminating triple-hypergeometric series correspond either to the 
semistretched isofactors of the second kind \cite{AlJ71} of Sp(4), or to 
the isofactors of the symmetric irreps of the orthogonal group SO($n$) 
in the canonical and semicanonical (tree type) bases. Our relation 
(\ref{s11ja})--(\ref{s11jc}) (which is significant within the framework 
of Sp(4) isofactors) is a triple-sum generalization of transformation 
formula (9) of \cite{LV-J01} for terminating $F_{1:1,1}^{1:2,2}$ 
Kamp\'{e} de F\'{e}riet series with a fixed single-integer non-positive 
parameter, restricting all summation parameters. (This restriction is 
hidden in equation (\ref{s11pa})--(\ref{s11pb}), rearranged for the aims 
of section 3.) Relations (\ref{s11ja})--(\ref{s11jc}), with intermediate 
formula (\ref{s11jb}), were derived using the transformation formulae 
\cite{Al00,LV-J01} of the double series, treated as the stretched $9j$ 
coefficients. The relation (\ref{iJpb})--(\ref{iJpd}) (important within 
the framework of SO($n$) isofactors) cannot be associated with any
transformation formula \cite{LV-J01} for terminating $F_{1:1,1}^{1:2,2}$
Kamp\'{e} de F\'{e}riet series with the same (single or double) 
parameters, restricting summation.

The wish may arise to prove identity (\ref{iJpb})--(\ref{iJpd}) by a 
direct transformation procedure, without using auxiliary rearrangement 
of section 2. Initially, the relation between (\ref{iJpb}) and analytical 
continuation of (\ref{s11jb}) (with the same three parameters 
restricting summation in the both cases) may be proved, using 
composition of transformation formulae of section III of \cite{LV-J01} 
for the double sum over $z_j$ and $z_k$ in (\ref{iJpb}) as terminating 
$F_{1:1,1}^{1:2,2}$ series into terminating $F_{0:2,2}^{1:2,2}$ series. 
Furthermore, the double sum version of relation between (\ref{iJpd}) 
(e.g. for $k_k=z_k=0$) and the result of previous step need to be 
considered. Both the sums over $z_l$ (for $\alpha _0$ and $\beta _0$ 
integers) can be recognized in the same Clebsch--Gordan coefficient of 
SU(2). Hence, the identity between two $_3F_2(1)$ series 
\cite{Sl66,GR90,RV-JR92} (for arbitrary $\alpha _0$ and $\beta _0$) 
induces the identity between the terminating double series. In its turn, 
inserting the latter one induces the identity between the terminating 
triple series, in which parameters restricting summation coincide only 
in part. Direct transformation of single $_3F_2(1)$ series 
\cite{Sl66,GR90,RV-JR92} is useless for proof of identities 
(\ref{s11ja})--(\ref{s11jc}) and (\ref{iJpb})--(\ref{iJpd}). 

Expressions (\ref{s11ja}) and (\ref{s11jc}) corresponding to special 
Sp(4) isofactors are summable or turn into the terminating Kamp\'{e} de 
F\'{e}riet \cite{K-F21,LV-J01} series $F_{2:1}^{2:2}$ for extreme 
basis states of Sp(4)$\supset $SU(2)$\times $SU(2). Alternatively, in 
accordance with (\ref{iGpc}) and (\ref{iJpd}), the expressions for 
special isofactors of SO($n$) and SU($n$) are summable in the case of the 
stretched couplings of the group representations and turn into the 
terminating Kamp\'{e} de F\'{e}riet series $F_{2:1}^{2:2}$ for the irreps 
of subgroups in a stretched situation, including the generic cases for 
restrictions SO$(n)\!\supset $SO($n-1$), 
SO$(n)\!\supset $SO$(n-2)\times $SO(2) and U$(n)\!\supset $U($n-1$). 
Taking into account the fact that the $F_{2:1}^{2:2}$ type series with 
five independent parameters also appeared as the denominator 
(normalization) functions of the SU(3) and $u_q(3)$ canonical tensor 
operators [46--48] %\cite{BLCC72,BLL85,LBL88} 
(cf (2.8) and section II of \cite{Al99}), the $q$-extension of relation 
(\ref{iJra})--(\ref{iJrb}) from the classical SU($n$) case may be 
suspected. 

In our next paper \cite{Al02b}, the fourfold \cite{Al87,HJu99} and 
(corrected) \cite{Al87} triple-sum expressions for the recoupling ($6l$) 
coefficients of symmetric irreps of SO($n$) will be rearranged into the 
double $F_{1:3}^{1:4}$ type series.

\appendix
\section{Special cases of triple-sums in $11j$ coefficients}

Rearranging (\ref{iJpst}) in inverse order that was used for transition
from (\ref{s11ja}) through (\ref{s11pa}) to (\ref{iJpb}), we derived an 
expression for special triple sum of the type (\ref{c11j5}) with the 
coinciding two first rows of the corresponding array, 
\bea
\fl \widetilde{{\bf S}}\left[ 
\begin{array}{cccc}
K_1 & j_1^1 & j_1^2 & j_1^3 \\ 
K_1 & j_1^1 & j_1^2 & j_1^3 \\ 
2K_1 & j^1 & j^2 & j^3
\end{array}
\right]  \nonumber \\
\lo= (-1)^{j_1^1+j_1^2+j_1^3-K_1-j^3}[1+(-1)^{j^1+j^2+j^3-2K_1}]
2^{j^1+j^2-2K_1-1}(2j^3-1)!j^3!  \nonumber \\
\times \prod_{a=1}^{2}\frac{(2j_1^a+j^a+1)!(j^a!)^2}{(2j^a+1)!
(2j_1^a-j^a)!}\sum_{x_1,x_2,z_3}\binom{j_1^3-\frac 12(j^3+\delta _3)
+x_3}{x_3}  \nonumber \\
\times \frac{(-1)^{x_3}\left( -j_1^3-(j^3+\delta _3)/2\right) _{x_3}}{%
(-j^3+1/2)_{x_3}}\binom{\frac 12(\delta _1+\delta _2-\delta _3)}{K_1
+j^3-\sum_{a=1}^3(\frac 12j^a-x_a)}  \nonumber \\
\times \prod_{a=1}^{2}\frac{\left( j_1^a+(j^a+\delta _a)/2
+1\right) _{x_a}}{(j^a+3/2)_{x_a}}\binom{j_1^a-\frac 12(j^a
+\delta _a)}{x_a},  \label{s11je2}
\eea
with $\delta _i=0$ or 1, so that $j_1^i-(j^i+\delta _i)/2$ 
($i=1,2,3 $) are integers.

Furthermore, similarly rearranging (\ref{iJpsr}), we derived an 
expression for special triple-sum $\widetilde{\cal S}[\cdot \cdot \cdot ]$
of the type (\ref{c11j5}) with the coinciding two first rows 
and $j_1^1=j_1^2,\;j_1^3=K_1$: 
\bea
\fl \widetilde{{\bf S}}\left[ \begin{array}{cccc}
j_1^3 & j_1^1 & j_1^1 & j_1^3 \\ 
j_1^3 & j_1^1 & j_1^1 & j_1^3 \\ 
2j_1^3 & j^1 & j^2 & j^3
\end{array} \right]  \nonumber \\
\lo= \frac{(2j_1^3-2j_1^1)!\Gamma (1/2)\Gamma \left( (j^1+j^2+j^3+1)/2
-j_1^3\right) \prod_{a=1}^3j^a!}{2^{4j_1^3+3}\prod_{a=1}^3\Gamma \left( %
j_1^3+(j^1+j^2+j^3+3)/2-j^a\right) }  \nonumber \\
\times \sum_{s}\frac{(2j_1^1+j^1+1)!(2j_1^1+j^2+1)!(2j_1^3+j^3+1)!}{s!
(2j_1^3-2j_1^1-s)!\left( j_1^3+(j^1-j^2-j^3)/2-s\right) !}  \nonumber \\
\times \frac{[1+(-1)^{j^1+j^2+j^3-2j_1^3}](-1)^{2j_1^1+j_1^3
-(j^1+j^2+j^3)/2}}{\left( j_1^3+(j^2-j^3-j^1)/2-s\right) !
\left( 2j_1^1-j_1^3+(j^3-j^1-j^2)/2+s\right) !}  \nonumber \\
\times \frac{(2j_1^1+1/2)_s\Gamma (2j_1^3+3/2-s)}{\left( 2j_1^1
-j_1^3+(j^1+j^2+j^3)/2+s+1\right) !}.  \label{s11je1}
\eea
Although this sum again corresponds to the balanced (Saalsch\"{u}tzian) 
$_4F_3(1)$ type series \cite{Sl66,GR90}, it is not alternating (since 
it includes even numbers of gamma functions or factorials in numerator 
and denominator) and cannot be associated with the $6j$ coefficients of 
SU(2). Note, that the summable case of (\ref{s11je1}) with 
$j_1^3=j_1^1$ corresponds to (41) of \cite{AlJ71}.

\section{On elementary proof of (\ref{iJpd})}

Expression (\ref{iJpc}) for the integrals involving the product of three 
Jacobi polynomials may be rearranged straightforwardly, without any 
allusion to special isofactors of Sp(4). For this purpose we apply the 
symmetry relation (\ref{iJp})--(\ref{iJpt}) (i.e.\ interchange 
$\alpha _a$ and $\beta _a$, $a=0,1,2,3$) to (\ref{iJpc}). When 
$\alpha _0$ and $\beta _0$ are integers, the $_3F_2(1)$ type sums 
over $z_i$ in modified expressions (\ref{iJpc}) and (\ref{iJpd}) 
correspond to the CG coefficients of SU(2) with the equivalent Regge 
$3\times 3$ symbols
\begin{equation}
\left\| \begin{array}{ccc}
k_i & k_i+\alpha _i+\beta _i & p_i-z_j-z_k \\ 
p_i^{\prime }\!+\!\beta  _i\!+\!k_j\!+\!k_k\!-\!z_j\!-\!z_k & 
p_i^{\prime \prime } & k_i+\alpha _i \\
p_i^{\prime \prime }+\alpha _i & 
p_i^{\prime }\!+\!k_j\!+\!k_k\!-\!z_j\!-\!z_k & k_i+\beta _i
\end{array} \right\|  \label{aRsc}
\end{equation}
and 
\begin{equation}
\left\| \begin{array}{ccc}
p_i-z_j-z_k & k_i & k_i+\alpha _i+\beta _i \\ 
k_i+\beta _i & p_i^{\prime \prime }+\alpha _i & 
p_i^{\prime }\!+\!k_j\!+\!k_k\!-\!z_j\!-\!z_k \\
k_i+\alpha _i & p_i^{\prime }\!+\!\beta _i\!+\!k_j\!+\!k_k\!-\!z_j\!-\!z_k 
& p_3^{\prime \prime }
\end{array} \right\| , \label{aRsd}
\end{equation}
expressed in the both cases by means of (15.1c) of Jucys and 
Bandzaitis \cite{JB77} [see also (7) of section 8.2 of \cite{VMK88}], but 
with hidden triangular conditions in the first case. For possible 
non-integer values of $\alpha _0$ and/or $\beta _0$, the doubts as to the 
equivalence of these finite $_3F_2(1)$ series may be caused by the 
absence of mutually coinciding integer parameters [$k_i$ in (\ref{iJpc}) 
and $\min (p_i-z_j-z_k,p_i^{\prime \prime })$ as triangular conditions in 
(\ref{iJpd}), respectively] restricting summation over $z_i$, unless 
equation (15.1d) of \cite{JB77} (together with possible inversion of 
summation) is used for the CG coefficient of SU(2) with Regge symbol 
(\ref{aRsd}). Note that the proof of relation between the corresponding 
finite $_3F_2(1)$ series in (\ref{iJpc}) and (\ref{iJpd}) based on 
composition of Thomae's transformation formulae (see \cite{Sl66,GR90}) 
or their Whipple's specifications for single restricting parameter 
(see \cite{Sl66,RV-JR92}) is rather complicated.

\end{document}